\documentclass[aps,prb,twocolumn,showpacs,preprintnumbers,superscriptaddress,amsmath,amssymb]{revtex4-1}

\usepackage[utf8]{inputenc} % symbols for accents
\usepackage[english]{babel} % language used
\usepackage{graphicx}% Include figure files
\usepackage{amssymb,amsmath} % math tools
\usepackage[hidelinks]{hyperref} % hyperlink to references
\usepackage{epstopdf}
\usepackage{nicefrac}
\usepackage{dcolumn}% Align table columns on decimal point
\usepackage{bm}% bold math
\usepackage{epsfig}
\usepackage{color}
\usepackage{siunitx} % SI units symbols
\usepackage{multirow}
\usepackage{physics}
\hypersetup{
	colorlinks,
	citecolor=blue,
	linkcolor=blue,
	urlcolor=blue,
}

\begin{document}
	
	\title{Crystalline and magnetic structure of Ba$_2$CuO$_{3+\delta}$ investigated by x-ray absorption spectroscopy and resonant inelastic x-ray scattering}
	
	\author{Roberto Fumagalli}
	\affiliation{Dipartimento di Fisica, Politecnico di Milano, Piazza Leonardo da Vinci 32, I-20133 Milano, Italy}
	
	\author{Abhishek Nag}
	\affiliation{Diamond Light Source, Harwell Campus, Didcot OX11 0DE, United Kingdom}
	
	\author{Stefano Agrestini}
	\altaffiliation{Present address: Diamond Light Source, Harwell Campus, Didcot OX11 0DE, United Kingdom}
	\affiliation{Max Planck Institute for Chemical Physics of Solids, Nöthnitzerstr. 40, 01187 Dresden, Germany}
	
	\author{Mirian Garcia-Fernandez}
	\affiliation{Diamond Light Source, Harwell Campus, Didcot OX11 0DE, United Kingdom}
	
	\author{Andrew C. Walters}
	\affiliation{Diamond Light Source, Harwell Campus, Didcot OX11 0DE, United Kingdom}
	
	\author{Davide Betto}
	\altaffiliation{Present address: Max-Planck-Institut f\"{u}r Festk\"{o}rperforschung, Heisenbergstr. 1, 70569 Stuttgart, Germany}
	\affiliation{ESRF -- The European Synchrotron, 71 Avenue des Martyrs, CS 40220, F-38043, Grenoble, France}
	
	\author{Nicholas B. Brookes}
	\affiliation{ESRF -- The European Synchrotron, 71 Avenue des Martyrs, CS 40220, F-38043, Grenoble, France}
	
	%\author{Z. Hu}
	%\affiliation{Max Planck Institute for Chemical Physics of Solids, Nöthnitzerstr. 40, 01187 Dresden, Germany}
	
	\author{Lucio Braicovich}
	\affiliation{Dipartimento di Fisica, Politecnico di Milano, Piazza Leonardo da Vinci 32, I-20133 Milano, Italy}
	\affiliation{ESRF -- The European Synchrotron, 71 Avenue des Martyrs, CS 40220, F-38043, Grenoble, France}
	
	\author{Ke-Jin Zhou}
	\email{kejin.zhou@diamond.ac.uk}
	\affiliation{Diamond Light Source, Harwell Campus, Didcot OX11 0DE, United Kingdom}
	
	\author{Giacomo Ghiringhelli}
	\email{giacomo.ghiringhelli@polimi.it}
	\affiliation{Dipartimento di Fisica, Politecnico di Milano, Piazza Leonardo da Vinci 32, I-20133 Milano, Italy}
	\affiliation{CNR-SPIN, Dipartimento di Fisica, Politecnico di Milano, 20133 Milano, Italy}
	
	\author{Marco Moretti Sala}
	\email{marco.moretti@polimi.it}
	\affiliation{Dipartimento di Fisica, Politecnico di Milano, Piazza Leonardo da Vinci 32, I-20133 Milano, Italy}
	
	\date{\today}
	
	\begin{abstract}
		
		Motivated by the recent synthesis of Ba$_2$CuO$_{3+\delta}$ (BCO), a high temperature superconducting cuprate with putative $d_{3z^2-r^2}$ ground state symmetry, we investigated its electronic structure by means of Cu $L_3$ x-ray absorption (XAS) and resonant inelastic x-ray scattering (RIXS) at the Cu $L_3$ edge on a polycrystalline sample. We show that the XAS profile of BCO is characterised by two peaks associated to inequivalent Cu sites, and that its RIXS response features a single, sharp peak associated to crystal-field excitations. We argue that these observations are only partially compatible with the previously proposed crystal structure of BCO. Based on our spectroscopic results and on previously published powder diffraction measurements, we propose a crystalline structure characterized by two inequivalent Cu sites located at alternated planes along the $c$ axis: nominally trivalent Cu(1) belonging to very short Cu-O chains, and divalent Cu(2) in the oxygen deficient CuO$_ {1.5}$ planes. We also analyze the low-energy region of the RIXS spectra to estimate the magnitude of the magnetic interactions in BCO and find that in-plane nearest neighbor superexchange exceeds 120~meV, similarly to that of other layered cuprates. Although these results do not support the pure  $d_{3z^2-r^2}$ ground state scenario, they hint at a significant departure from the common quasi-2D electronic structure of superconducting cuprates of pure $d_{x^2-y^2}$ symmetry.  
		
	\end{abstract}
	
	\maketitle
	
	\section{Introduction}
	
	Since the discovery of high temperature superconductivity in cuprates\cite{BednorzMuller1986}, many efforts have been spent to provide a conclusive and generally accepted explanation of this phenomenon. Unfortunately, results obtained so far are only partially satisfactory \cite{KeimerKivelsonReviewHTS}. 
	
	The discovery of novel families of unconventional superconductors is extremely important to help distinguishing indispensable and unnecessary, or even competing, ingredients for superconductivity. In this respect, the recent synthesis of Nd$_{0.8}$Sr$_{0.2}$NiO$_2$, an infinite-layer superconducting nickelate, is extremely welcome\cite{Li2019_NickelateSC} and, despite dissimilarities with superconducting cuprates, the paradigm that a 3$d^9$ ground state with $d_{x^2-y^2}$ orbital symmetry of the hole is key to achieve high temperature superconductivity is further strengthened \cite{Wu2019_arxiv,Sakakibara2019_arxiv,Botana2019_arxiv,Jiang2019_arxiv}. Li \emph{et al.}, however, challenged this belief by synthesising Ba$_2$CuO$_{3+\delta}$, a high temperature superconducting cuprate with putative $d_{3z^2-r^2}$ ground state symmetry \cite{Li_2019}.
	
	Most cuprates are layered materials, with superconducting CuO$_2$ planes intercalated by blocking layers that act as charge reservoirs. In their antiferromagnetic, insulating parent compounds the ground state has predominant $d_{x^2-y^2}$ symmetry\cite{Chen_1992} due to the sign of the tetragonal crystal field acting on the Cu$^{2+}$-derived electronic states. Moreover, it has recently been suggested that the larger the energy splitting between the $d_{x^2-y^2}$ and the $d_{3z^2-r^2}$ states, the higher the superconducting critical temperature\cite{Peng_NatPhys}. Ba$_2$CuO$_{3+\delta}$ (BCO) is supposedly isostructural to the prototypical (La$_{2-x}$,Ba$_x$)CuO$_4$ high-temperature cuprate superconductors, which adopt a K$_2$NiF$_4$-type structure, where a sizeable elongation of the CuO$_6$ octahedra in the $c$ axis direction typically stabilizes a ground state with  $d_{x^2-y^2}$ symmetry. In BCO, instead, it was suggested that Cu-O bonds might be shorter out of plane than in plane, leading to a ground state with predominantly $d_{3z^2-r^2}$ character. This scenario is intriguing, because it undermines the relevance of the $d_{x^2-y^2}$ ground state for high temperature superconductivity and, therefore, motivates further investigations. At a theoretical level, a mechanism for the pairing that heavily relies on the significant weight of both $d_{x^2-y^2}$ and $d_{3z^2-r^2}$ orbitals at the Fermi energy has been recently proposed\cite{Maier_PRB2019}. Experimentally, instead, little has been done besides the initial work of Li \emph{et al.} \cite{Li_2019}.
	
	Here we probe BCO by means of x-ray absorption spectroscopy (XAS) and resonant inelastic x-ray scattering (RIXS) at the Cu $L_3$-edge to set constraints on its crystal, magnetic and electronic structure. Experimental limitations due to the use of a polycrystalline sample prevented us from exploiting all the possibilities offered by these techniques; in particular, we could not investigate the angular and polarization dependence of XAS and RIXS cross-sections and determine the symmetry of the ground and excited states in BCO, as done in the past for other cuprates\cite{ddMoretti}. Nevertheless, we show  that our results are only partially compatible with the crystal structure previously proposed for BCO and propose an alternative, consistent with our and previous measurements.
	
	%The present paper is organized as follows: in Sec.~\ref{Exp_Methods} we introduce the experimental details; in Sec.~\ref{Results} we present our main findings, which will then be discussed in Sec.~\ref{Discussion} in comparison to other model cuprate systems.
	
	\section{Experimental methods}
	\label{Exp_Methods}
	
	Our powder sample of Ba$_2$CuO$_{3+\delta}$ ($\delta \approx 0.2$) is characterized by a superconducting critical temperature $T_c\approx70$ K and was synthesised according to the procedure described in Ref.~\onlinecite{Li_2019}. 
	
	XAS and RIXS spectra were collected at the I21 beam line of the Diamond Light Source (Didcot, UK). XAS measurements were carried out in total fluorescence yield mode. RIXS spectra were measured with an overall energy resolution of approximately 55~meV. The x-ray beam spot size was approximately $3\times20$ $\mu$m$^2$. The polarization of the incident photons was vertical in the laboratory reference frame and perpendicular to the scattering plane. The scattering angle $2\theta$ was kept fixed at 154$^\circ$. The polycrystalline sample was compressed in a pellet, mounted on the sample holder and then sealed in a vacuum suitcase in the protected atmosphere of a glovebox. The vacuum suitcase was brought to the beam line and connected onto the sample load-lock. In order to expose a fresh sample surface, the pellet was cleaved at a vacuum pressure lower than $10^{-7}$~mbar. All measurements were carried out at a temperature of 20 K and a base vacuum pressure lower than $5\times10^{-10}$~mbar. 
	\section{Results}
	\label{Results}
	
\begin{figure}[htbp!]
		\centering
		\includegraphics[width=1\columnwidth]{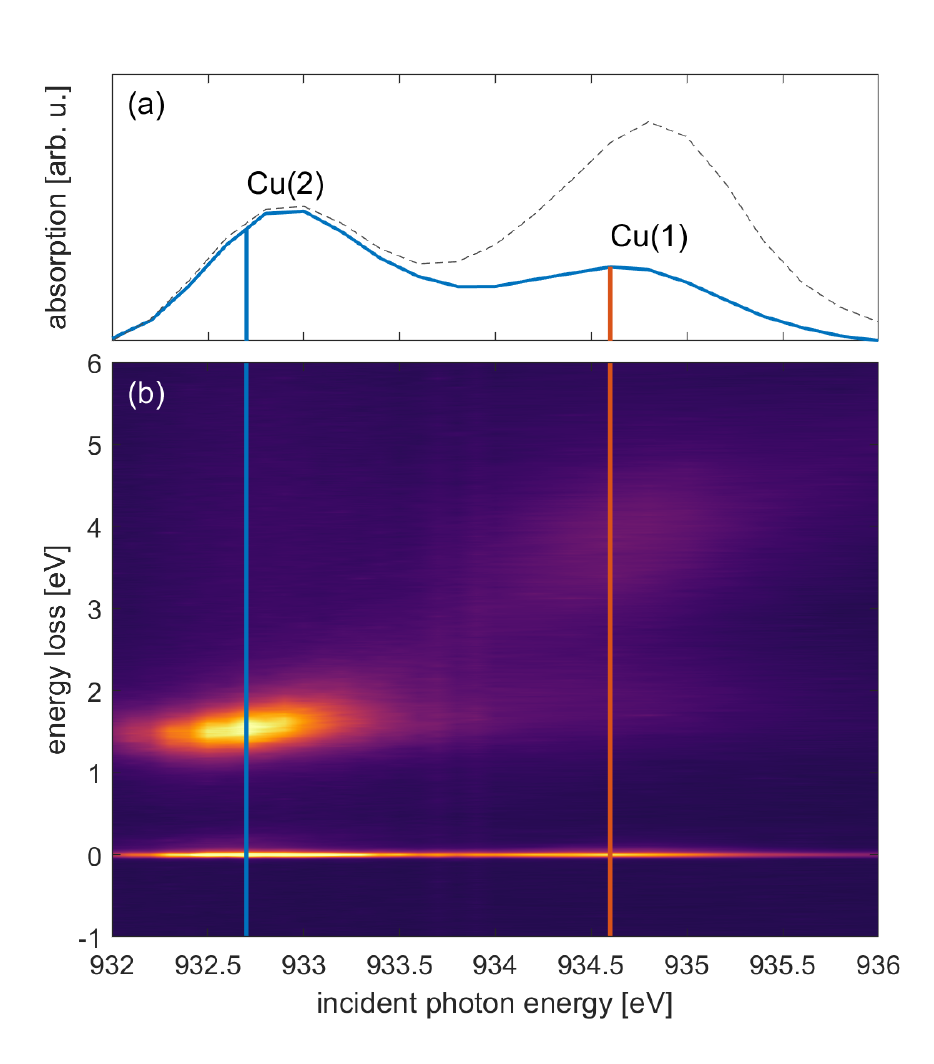}
		\caption{(a) Cu L$_3$ edge XAS and (b) RIXS map of powder BCO at 20 K. The vertical lines correspond to the cuts reported in Fig. \ref{fig_RIXS_cuts}. The dashed black line in panel (a) corresponds to the XAS profile acquired on a different region of the sample (see main text for further details).}
		\label{fig_RIXS_map}
	\end{figure}

	We report in Fig.~\ref{fig_RIXS_map}(a) the x-ray absorption profile of BCO across the Cu $L_3$ absorption edge. It shows two distinct features (to be discussed in the following) at approximately 932.9~eV and 934.6~eV, respectively.
	
	We collected RIXS spectra for incident photon energies in the energy window corresponding to the XAS spectrum and show the results in Fig.~\ref{fig_RIXS_map}(b). The strong, narrow feature at constant zero energy loss is the elastic line, while higher energy loss features correspond to inelastic excitations which depend on the incident photon energy. The intense peak at approximately 1.5~eV energy loss is maximised in the vicinity of the first peak in the XAS spectrum ($\sim 932.9$~eV) and its position is rather constant as the incident photon energy is scanned. In agreement with a number of RIXS studies on cuprates\cite{Ament_review,ddMoretti}, these features are assigned to crystal field (or $dd$) and charge-transfer excitations. A broader and weaker feature is observed at higher energy losses: it is strongest at incident photon energies corresponding to the second peak in the XAS spectrum and partially displays a fluorescence-like behaviour as seen, for example, in YBa$_2$Cu$_3$O$_{6.99}$ (see Supplemental Materials of Ref.~\onlinecite{Minola_PRL2015}). 
	
\begin{figure}[htbp!]
		\centering
		\includegraphics[width=1\columnwidth]{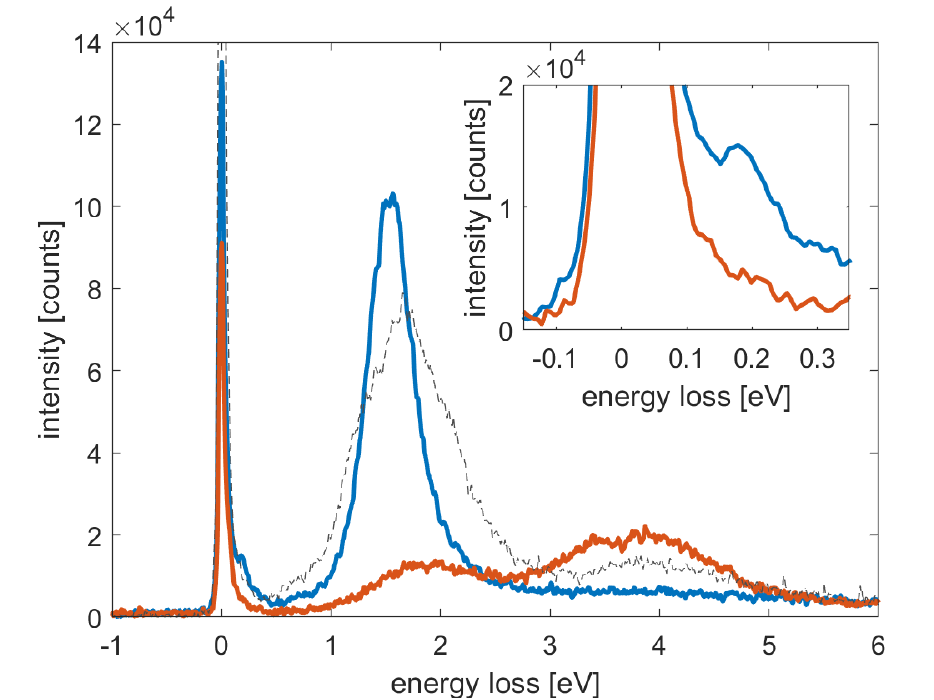}
		\caption{RIXS spectra of polycrystalline BCO measured at incident photon energies of 932.7 (solid blue, dashed black) and 934.6~eV (solid red line), corresponding to the vertical lines in Fig.~\ref{fig_RIXS_map}, respectively. The dashed black curve is the RIXS spectum acquired on the region of the sample corresponding to the dashed black XAS profile of Fig.~\ref{fig_RIXS_map}(a). The inset shows a close-up view of the low energy loss window.}
		\label{fig_RIXS_cuts}
	\end{figure}

	A closer inspection of the RIXS data is provided in Fig.~\ref{fig_RIXS_cuts}, which shows two vertical cuts of the RIXS map of Fig.~\ref{fig_RIXS_map}(b). The overall shape of the spectra is markedly different: in particular, the RIXS spectrum taken at low incident photon energy (blue) shows a single, narrow peak at 1.6~eV energy loss, whereas the RIXS spectrum taken at high incident photon energy (red line) is characterized by two, broad features at approximately 2 and 4~eV. In addition, we highlight in the inset the presence of a feature at approximately 0.2~eV, which resonates at the lower energy peak in the XAS spectrum.
	
\begin{figure}[htbp!]
		\centering
		\includegraphics[width=1\columnwidth]{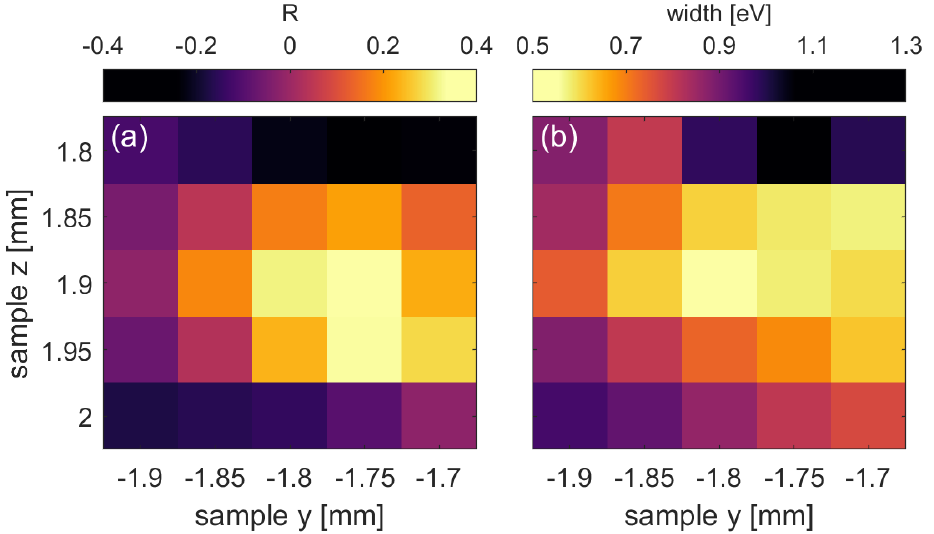}
		\caption{Spatial dependence of (a) the ratio between the two peaks in the XAS spectrum of Fig. \ref{fig_RIXS_map}(a) and (b) the width of the 1.6~eV excitation in the RIXS spectrum of Fig. \ref{fig_RIXS_cuts}.}
		\label{fig_microscopy}
	\end{figure}

	Before moving to the discussion of the experimental results, we note that the polycrystalline BCO sample was found to be rather inhomogeneous as observed with both XAS and RIXS. The black dashed lines in Figs. \ref{fig_RIXS_map}(a) and \ref{fig_RIXS_cuts} correspond to XAS and RIXS spectra, respectively, taken on a different area of the sample. The inhomogeneity is demonstrated in the maps of Fig. \ref{fig_microscopy}: in panel (a) we show the spatial dependence of the relative intensity change of the two XAS peaks through the ratio 
	\begin{equation}
	R = \frac{I(932.7\ \mathrm{eV})-I(934.6\ \mathrm{eV})}{I(932.7\ \mathrm{eV})+I(934.6\ \mathrm{eV})},
	\end{equation}  
	where $I(X)$ is the XAS intensity at energy $X$; in panel (b) we show the spatial dependence of the full width at half maximum of crystal-field excitations in the RIXS spectra. The two maps show a clear correlation, and the first peak in the XAS is larger in the part of the sample where the crystal field transition in RIXS is sharper. The length scale of the spatial variations amounts to a few tens of microns We are aware of the fact that these samples are particularly sensitive to air exposure: different grains might have been chemically stabilized differently, leading to spatial variations of doping and composition. In agreement with literature\cite{Li_2019}, we here focus out attention on the region of the sample where the second XAS peak is weaker. Moreover we use the RIXS spectra excited at the first XAS peak, thus selecting mostly Cu$^{2+}$ sites in the ``good'' phase.
	
	\section{Discussion}
	\label{Discussion}
	
\begin{figure}[htbp!]
		\centering
		\includegraphics[width=1\columnwidth]{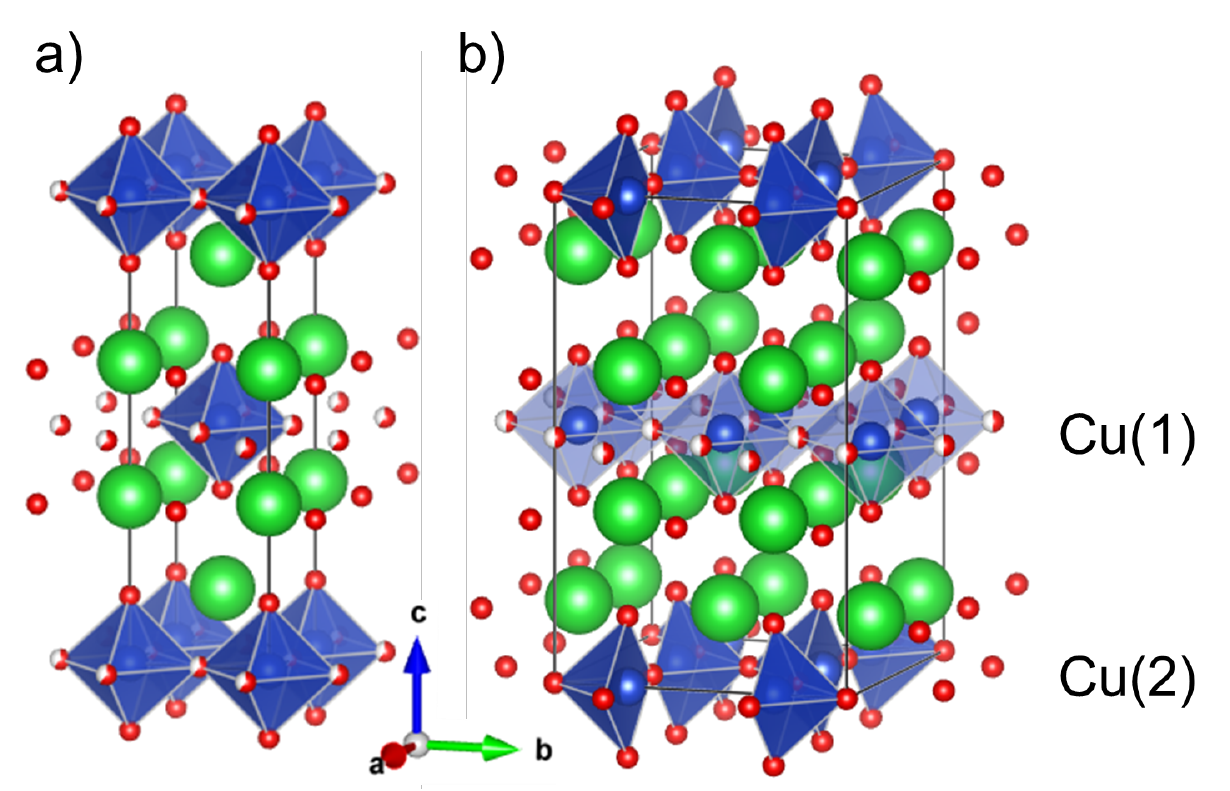}
		\caption{Crystal structure of BCO as reported in Ref.~\onlinecite{Li_2019} (a) and as proposed in this work (b). Green, blue and (partially filled) red balls represent Ba, Cu and (partially occupied) O sites.}
		\label{fig_xtal_structures}
	\end{figure}

	We start the discussion by commenting on the Cu L$_3$ edge XAS spectrum of BCO: irrespective of its detailed interpretation, the presence of two distinct features suggests the presence of two, non-equivalent Cu sites in the system. As established in the literature on cuprates, the first and second XAS peak in Fig.~\ref{fig_RIXS_map}(a) can be associated to nominal Cu$^{2+}$ and Cu$^{3+}$ species, respectively\cite{Hawthorn_2011,Minola_PRL2015}. However, in the structure previously proposed for BCO all the Cu sites are considered to be equivalent, as summarized in Fig.~\ref{fig_xtal_structures}(a) and Tab.~\ref{tab_BCO}. The double XAS peak, in that scenario, would be possible only in the (unlikely) case of a strong charge discommensuration. Here we have explored other scenarios consistent both with the powder diffraction patterns shown in Ref. \onlinecite{Li_2019} and with the present XAS (and RIXS) data that suggest the existence of two inequivalent Cu sites, in analogy to YBCO. To keep the analogy with YBCO we label Cu(1) and Cu(2) the Cu$^{3+}$ and Cu$^{2+}$ site respectively. Based on the maps of Fig.~\ref{fig_RIXS_map} and the spectra of Fig.~\ref{fig_RIXS_cuts} we will concentrate on RIXS spectra excited at 932.7~eV, which correspond to selecting Cu(2) sites only.
	
\begin{figure}[htbp!]
		\centering
		\includegraphics[width=1\columnwidth]{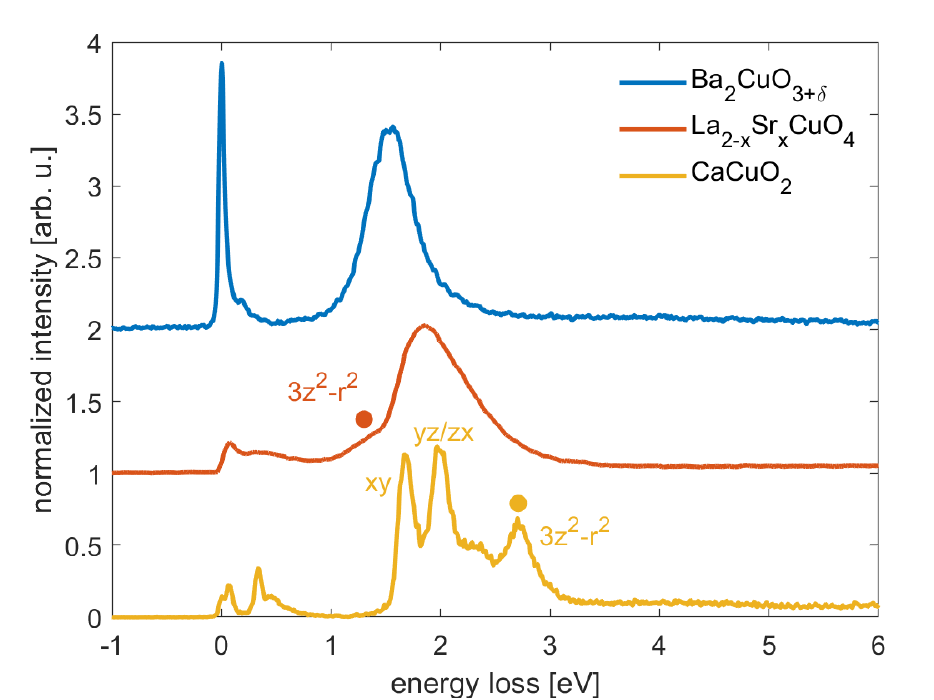}
		\caption{RIXS spectra of powder BCO (as in Fig.~\ref{fig_RIXS_cuts}, blue), compared to single-crystalline LSCO\cite{Dean_2013} (red) and CCO (yellow line) reference systems, normalized to the spectral weight of the $dd$ excitations.}
		\label{fig_RIXS_comparison}
	\end{figure}

	In Fig.~\ref{fig_RIXS_comparison}, we compare the crystal field excitations in superconducting BCO and two prototypical cuprate systems, i.e., the single-layer superconductor La$_{2-x}$Sr$_x$CuO$_4$ (LSCO, $x=0.15$)\cite{Dean_2013} and the infinite-layer insulator CaCuO$_2$ (CCO).
	
	Crystal field (or $dd$) excitations arise from the redistribution of electrons within the atomic $3d$ shell through the RIXS process. For a Cu$^{2+}$ ion, a single particle approach is justified and $dd$ excitations simply correspond to transitions between crystal field split states\cite{Ament_review,ddMoretti}. The symmetry and energy of $3d$ states is thus determined by the geometrical arrangement of the nearest neighbour oxygen ions around the central Cu$^{2+}$ ion, whereas the Cu-O bond distances set the magnitude of the splitting. The scattering geometry and polarization dependence of the RIXS cross-sections for $dd$ excitations have been successfully used to determine the symmetry of the ground and excited states of several cuprates \cite{ddMoretti,Kang_2019,Fumagalli_2019}. In particular, the assignment of $dd$ excitations in CCO is as follows: the peaks at approximately 1.7, 2.0 and 2.7~eV correspond to transitions from the $d_{x^2-y^2}$ ground state to the $d_{xy}$, $d_{yz/zx}$ and $d_{3z^2-r^2}$ excited states, respectively; in the case of LSCO, doping causes an overall broadening of the spectrum, but one can still appreciate that the energies of the $d_{xy}$ and $d_{yz/zx}$ transitions are similar to those in CCO, whereas the $d_{3z^2-r^2}$ excited state is found around 1.4~eV, in analogy to the single-layer insulating parent compound La$_2$CuO$_4$. We note that the energy of the $d_{xy}$ excited state is mainly dictated by the in-plane Cu-O bond distance, which is similar for LSCO (1.91~\AA)\cite{Manske_2004} and CCO (1.93~\AA)\cite{Siegrist_1988}. Instead, the energy of the $d_{3z^2-r^2}$ excited state is mostly determined by the out-of-plane coordination of Cu\cite{Hozoi_2011}: in LSCO, the out-of-plane Cu-O bond distance is 2.46~\AA, while the infinite-layer structure of CCO features CuO$_4$ plaquettes with no apical oxygens (also shown in Fig.~\ref{fig_RIXS_comparison}), which pushes the $d_{3z^2-r^2}$ excited state to higher energies. 
	
	%Unfortunately, in the case of a powder sample the scattering geometry and polarization dependence of the RIXS cross-sections is largely washed out and, as a consequence, it cannot be exploited to infer about the ground and excited state symmetry. Still, there is a wealth of information that can be extracted from the experimental data. (GIACOMO SUGGERISCE DI TOGLIERE QUESTA PARTE, E' UNA RIPETIZIONE CHE TAGLIA IL DISCORSO) 
	
	\begin{figure}[htbp!]
		\centering
		\includegraphics[width=1\columnwidth]{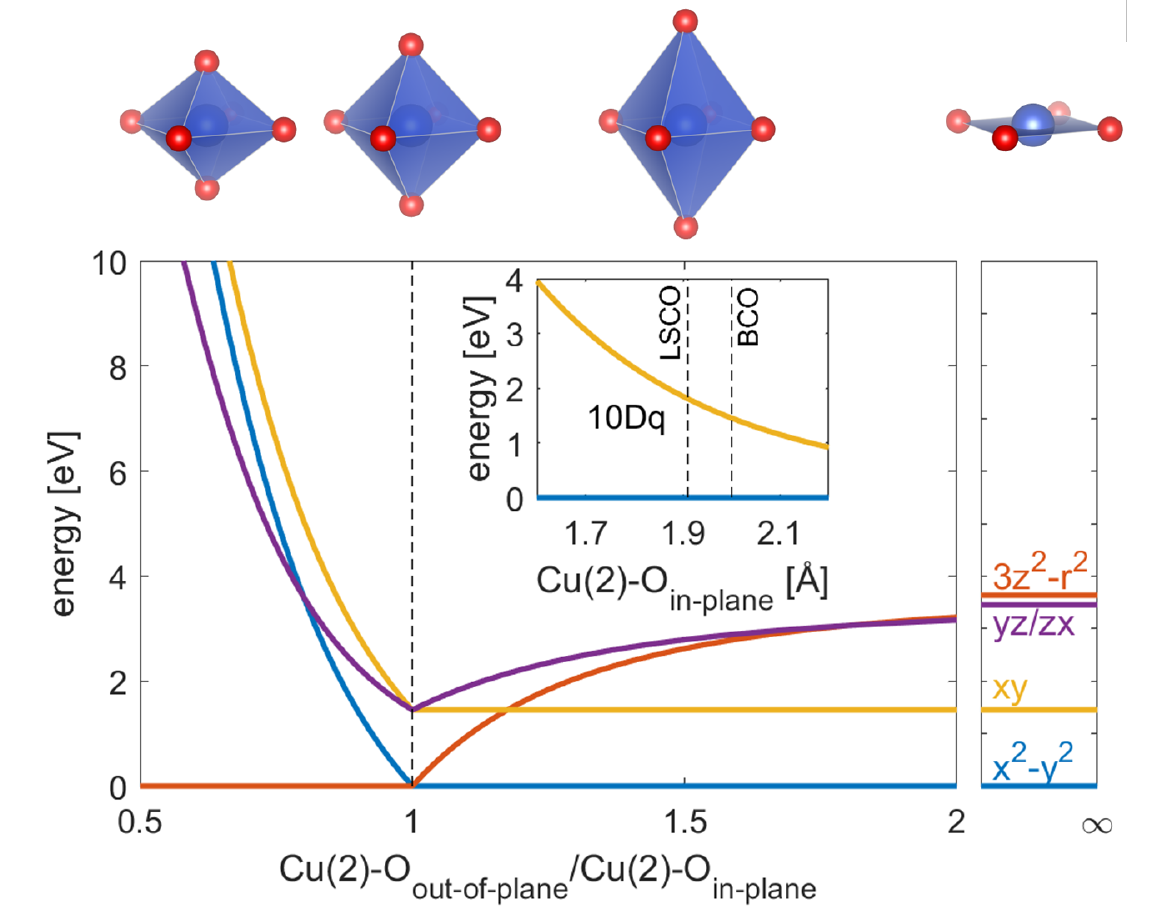}
		\caption{Energy of the Cu 3$d$ states as a function of the out-of-plane/in-plane Cu-ligand distance\cite{Bersuker}. In the inset we show the $10Dq$ energy as a function of the in plane Cu(2)-O bonds.}
		\label{fig_CF}
	\end{figure}
	
	The RIXS spectrum of BCO is remarkably different from those of CCO and LSCO, and of all the layered cuprates studied so far. In particular, Fig.~\ref{fig_RIXS_comparison} suggests three main observations: $dd$ excitations in BCO are i) lower in energy than any other cuprate system\cite{Hozoi_2011,ddMoretti}, ii) sharper and iii) more symmetric than in LSCO. 
	
	\figurename~\ref{fig_CF} shows the energy of the Cu $3d$ states as a function of the ratio between the out-of-plane and in-plane Cu(2)-O distance obtained from simple crystal field considerations reported in Ref.~\onlinecite{Bersuker}: we consider an hydrogen-like model for Cu$^{2+}$ surrounded by six $-2e$ point charges at the corners of  an octahedron and an effective nuclear charge $Z_\mathrm{eff}=6.7$ such that $10Dq = 1.8$~eV for Cu(2)-O$_\textrm{in-plane}= 1.91$ \AA\ (as in La$_2$CuO$_4$ (LCO)\cite{ddMoretti}) and provides a qualitative picture of the energy level splitting. In the case of an  octahedron compressed along the $z$ axis, the ground state has the $d_{3z^2-r^2}$ symmetry of the only $3d$ hole of Cu$^{2+}$, whereas for an elongated octahedron the ground state is characterized by a $d_{x^2-y^2}$  symmetry. In both cases, the energy of the other 3$d$ states is strongly affected by the effective Cu(2)-O$_{\textrm{out-of-plane}}$/Cu(2)-O$_{\textrm{in-plane}}$ ratio. The crystal field model suggests that the energy of the $d_{3z^2-r^2}$ and the two-fold degenerate $d_{xz/yz}$ states is mostly influenced by the Cu(2)-O$_\textrm{out-of-plane}$ distance, while 10Dq, i.e. the energy difference between the $d_{x^2-y^2}$ and $d_{xy}$ orbitals, is mostly dictated by the Cu(2)-O$_\textrm{in-plane}$ distance, as reported in the inset of \figurename~\ref{fig_CF}. The in-plane Cu(2)-O bond distance is considerably larger in BCO (2.00 \AA\cite{Li_2019arxiv}) than in LSCO and CCO, thus explaining the overall lower energy of $dd$ excitations in BCO than in the other cuprates). Indeed, the decrease of $10Dq$ with the Cu(2)-O$_\textrm{in-plane}$ distance of \figurename~\ref{fig_CF} can be fit to a power decay with exponent -4.5; one can then rescale $10Dq$ of LCO with the lattice parameter of BCO and obtain $1.8\ \mathrm{eV}\times(2.00\ \mathrm{\AA}/1.91\ \mathrm{\AA})^{-4.5} \approx 1.5\ \mathrm{eV}$, in agreement with our observations. %However, we will argue in the following that, besides the XAS, also ii) and iii) cannot be accounted for by assuming the crystal structure previously proposed for BCO\cite{Li_2019,Liu_PhysRevMaterials2019}. TOGLIEREI QUESTA FRASE%
	
	\begin{table}
		\centering
		\begin{tabular}{c c c c c c c c} 
			\hline
			\multicolumn{8}{l}{Space group : $I4/mmm$ (\# 139)}\\
			\multicolumn{8}{l}{$a$ = $b$ = 4.003 \AA, $c$ = 12.942 \AA}\\
			\hline		
			label & symbol & mult. & Wyckoff & $x$ & $y$ & $z$ & occupancy\\ 
			\hline\hline
			Ba(1) & Ba & 4 & e & 0.000 & 0.000 & 0.356 & 1.000\\ 
			Cu(1) & Cu & 2 & a & 0.000 & 0.000 & 0.000 & 1.000\\  
			O(1)  & O  & 4 & e & 0.000 & 0.000 & 0.144 & 1.000\\ 
			O(2)  & O  & 4 & c & 0.000 & 0.500 & 0.000 & 0.625\\ 
			\hline
		\end{tabular}
		\caption{Principal parameters for the crystal structure of BCO ($\delta = 0.25$) for the crystal structure reported in Ref.~\onlinecite{Le_2019_arxiv} and shown in Fig.~\ref{fig_xtal_structures}(a).}
		\label{tab_BCO}
	\end{table}
	
	Beyond the information on the average crystal field energy and Cu-O distances, RIXS spectra can be used to learn more about the crystalline structure of BCO. For the sake of simplicity in the discussion, we consider a nominal $\delta = 0.25$, corresponding to the formula with integer stoichiometric coefficients Ba$_8$Cu$_4$O$_{13}$. In this case, the in-plane oxygens should have a fractional occupancy of $\frac{13}{16}$, i.e., the Cu ions are coordinated to 4.5 oxygens, on average, instead of 6 as in LCO. Supposing a random distribution of oxygen vacancies and ignoring, for a moment, the distinction between Cu(1) and Cu(2) sites, one would expect a variety of coordinations for Cu ions, with 3, 4, 5 and 6 oxygen ligands, and a consequential distribution of dd excitation energies. Indeed, it has been shown that disorder leads to broadening of the $dd$ excitation multiplet in RIXS \cite{Fumagalli_2019}. Therefore,  the observation ii) of a very narrow crystal field excitation feature in the RIXS spectrum of BCO does not seem to be consistent with a highly disordered system. Moreover, the average coordination of Cu imposes that a large number of CuO$_4$ plaquettes should form in BCO, therefore contributing to the RIXS spectrum with an excitation at high energy losses ($\gtrsim 2.5$~eV), similarly to the $d_{3z^2-r^2}$ transition in CCO. The observation iii) of a very symmetric line shape excludes the presence of such a high energy peak, therefore suggesting that most of the Cu ions have a coordination number larger than 4, at least for the divalent sites Cu(2) probed by RIXS.
	
	Recently, Liu \textit{et al.} \cite{Liu_PhysRevMaterials2019} confirmed that oxygen vacancies prefer to reside in the planar rather than the apical sites and suggested alternative, more stable crystals structures; in particular, they considered a $2 \times 2 \times 1$ supercell and calculated the energies associated to 26 possible crystal structures with various arrangements of 6 oxygen vacancies (corresponding to the $\delta=0.25$ phase we are considering here). They found that the most favourable structures contain 6 and 4 oxygens within layers at fractional coordinates $z=0$ and $z=0.5$, respectively. Indeed, those structures imply three inequivalent Cu sites: one 6-fold coordinated at $z=0$, and two 4-fold coordinated at $z=0$ and $z=0.5$. To comply with our experimental evidence of 2 and not 3 Cu sites, we propose a modified structure, shown in Fig.~\ref{fig_xtal_structures}(b), that still respects the property of having 6 oxygens at $z=0$ and 4 at $z=0.5$ as in Ref.~\onlinecite{Liu_PhysRevMaterials2019}. Moreover, it features only two inequivalent Cu sites,  Cu(2)$^{2+}$ and Cu(1)$^{3+}$, within the $z=0$ and $z=0.5$ layers, respectively. And the Cu$^{2+}$ ions, contributing to the RIXS spectra of Fig.~\ref{fig_RIXS_comparison}, are 5-fold coordinated in pyramids, while Cu$^{3+}$ ions, silent for RIXS at 932.7~eV incident photon energy, are 4-fold coordinated in plaquettes. It is important to note that, in our model, the plaquettes in the $z=0.5$ layer do not form long chains but are equally and randomly distributed between the $ac$ and the $bc$ planes to respect the tetragonal symmetry of the structure. This condition is also needed in consideration of the XAS spectral shape. In fact, when the Cu-O-Cu bond angle is 180$^\circ$, like in chains of corner-shared CuO$_4$ plaquettes, the inter-plaquette electron hopping is maximized and non-local processes are able to screen the core hole in Cu$^{3+}$ L$_3$ XAS and the corresponding peak in the XAS spectrum becomes very broad and barely visible \cite{Chen_1992, Merz1998}. Instead, when the Cu-O-Cu bond angle is close to 90$^\circ$ the inter-plaquette hybridisation is strongly suppressed and non-local ligand electrons cannot take part in the core-hole screening, which results in a sharp and well defined Cu$^{3+}$ peak \cite{Hu2002}. The XAS of BCO showing two well defined peaks clearly agrees with the latter scenario and indicates that the plaquettes must be disordered, though all oriented perpendicular to the $ab$ plane. Tables~\ref{tab_BCO_b} reports additional information on the crystal structure of BCO that we are proposing here.
	
	\begin{table}
		\centering
		\begin{tabular}{c c c c c c c c} 
			\hline
			\multicolumn{8}{l}{Space group : $Cmmm$ (\# 65)}\\
			\multicolumn{8}{l}{$a$ = $b$ = 8.006 \AA, $c$ = 12.942 \AA}\\
			\hline
			label & symbol & mult. & Wyckoff & $x$ & $y$ & $z$ & occupancy\\ 
			\hline\hline
			Ba(1) & Ba & 8 & n & 0.250 & 0.000 & 0.144 & 1.000\\ 
			Ba(2) & Ba & 8 & o & 0.000 & 0.750 & 0.357 & 1.000\\ 
			Cu(1) & Cu & 4 & j & 0.250 & 0.000 & 0.500 & 1.000\\
			Cu(2) & Cu & 4 & g & 0.000 & 0.750 & 0.000 & 1.000\\  
			O(1)  & O  & 2 & a & 0.000 & 0.000 & 0.000 & 1.000\\   
			O(2)  & O  & 4 & e & 0.250 & 0.250 & 0.000 & 1.000\\   
			O(3)  & O  & 8 & o & 0.000 & 0.750 & 0.144 & 1.000\\ 
			O(4)  & O  & 8 & n & 0.250 & 0.000 & 0.357 & 1.000\\ 
			O(5)  & O  & 2 & d & 0.000 & 0.000 & 0.500 & 0.500\\   
			O(6)  & O  & 2 & c & 0.000 & 0.500 & 0.500 & 0.500\\   
			\hline
		\end{tabular}
		\caption{Principal parameters for the crystal structure of BCO ($\delta = 0.25$) as proposed in this work and shown in Fig.~\ref{fig_xtal_structures}(b).}
		\label{tab_BCO_b}
	\end{table}
	
	\begin{figure}[htbp!]
		\centering
		\includegraphics[width=1\columnwidth]{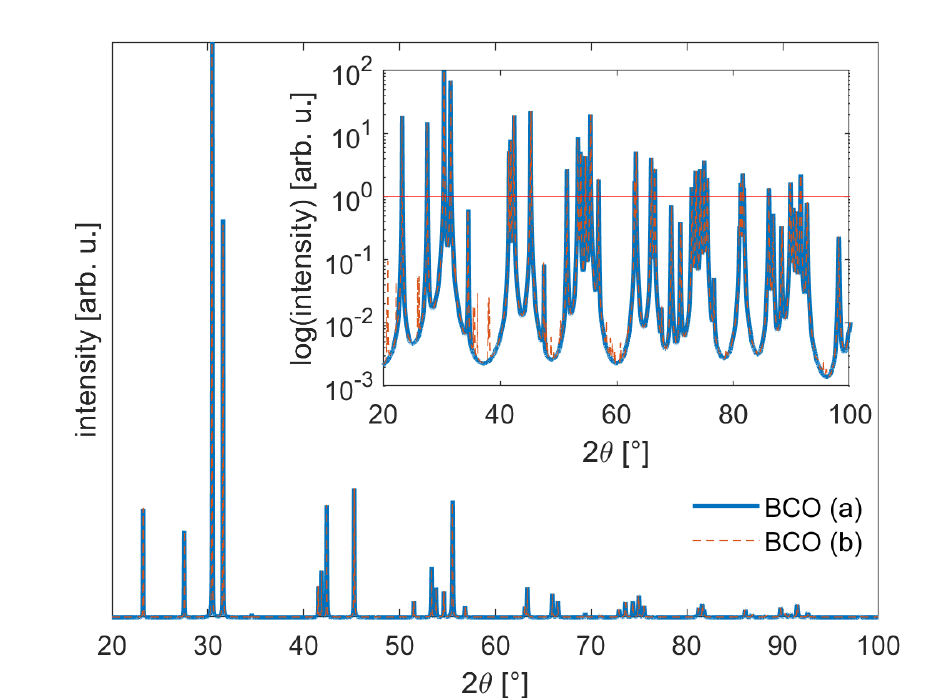}
		\caption{Simulated pXRD patterns for the crystal structures reported in Fig.~\ref{fig_xtal_structures}(a) and (b).}
		\label{fig_powder_XRD}
	\end{figure}
	
\begin{figure}[htbp!]
		\centering
		\includegraphics[width=1\columnwidth]{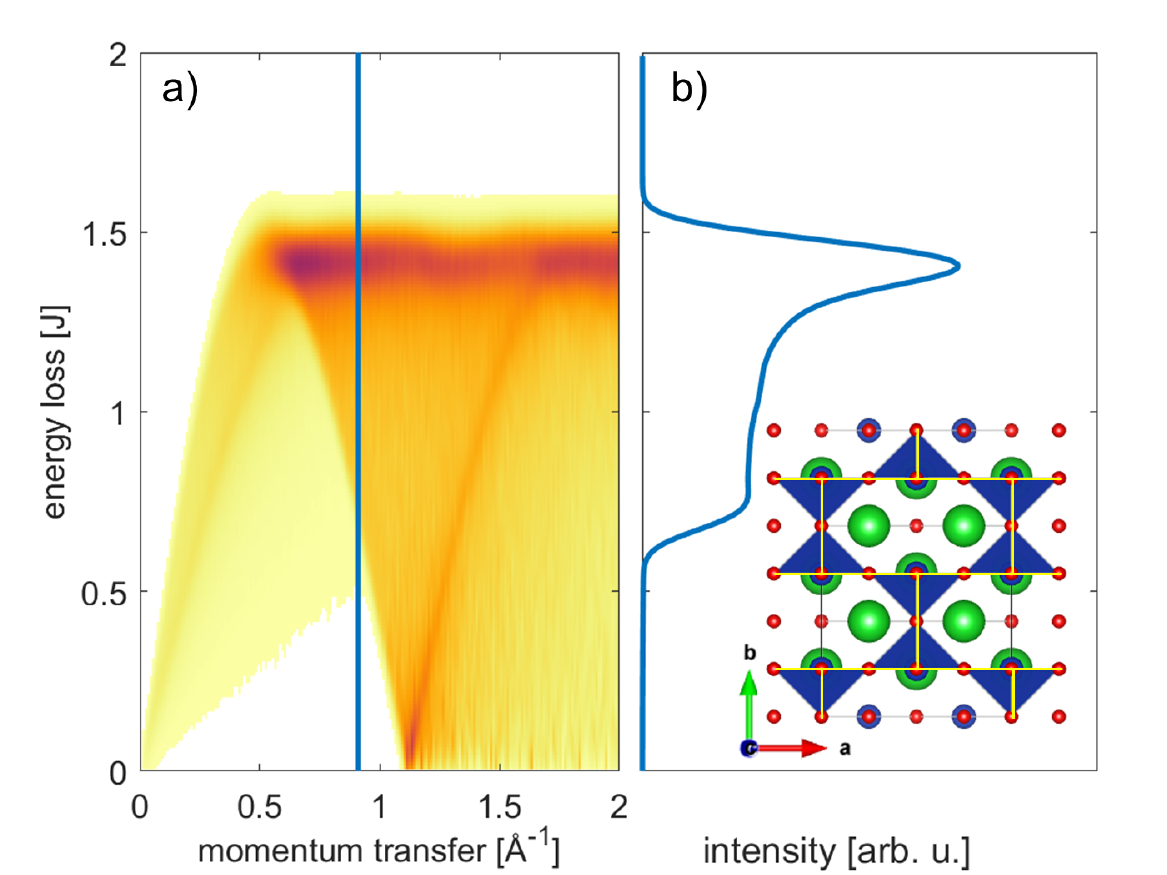}
		\caption{(a) Simulated momentum dependence of the magnetic structure factor and (b) its cut at 0.91 \AA$^{-1}$, corresponding to scattering angles of $\theta=154^{\circ}$ for powder BCO. The considered nearest-neighbours superxchange interactions can be seen from the $ab$-plane crystal structure of BCO, also shown.}
		\label{fig_RIXS_powder_sim}
	\end{figure}

	The crystal structure originally proposed for BCO was obtained by fitting its experimental powder x-ray diffraction (pXRD) pattern\cite{Li_2019}. The corresponding simulated pXRD pattern is reported in Fig.~\ref{fig_RIXS_powder_sim} as a solid blue line, which we use as a reference for comparison with the simulated pXRD pattern of the crystal structure we propose in this work: it turns out that the pXRD patterns of the crystal structures reported in~\ref{fig_xtal_structures}(a) and (b) largely overlap and are indistinguishable at a first sight. When the intensities are plotted on a logarithmic scale (inset), little discrepancies that become apparent are below the detectability limit of the pXRD experiment (red line), estimated from the (in)visible peaks in the experimental data \cite{Li_2019}. Therefore, we believe that the structure we propose here is as compatible as the one proposed originally \cite{Li_2019} with the currently available experimental pXRD pattern. 
	
	Having established that the crystal structure of Fig.~\ref{fig_xtal_structures}(b) is compatible with previous pXRD measurements, we now discuss implications on the electronic properties of BCO. We expect that the superconductivity takes place in the $z = 0$ layer, while the $z = 0.5$ layer plays the role of charge reservoir, in analogy to YBCO. Therefore, we would like to discuss in more detail the ground state symmetry of the Cu(2) ions. As shown in the inset of Fig.~\ref{fig_xtal_structures}(b) the Cu$^{2+}$ ions are coordinated in pyramids, having their base in the $ac$ plane. In a regular pyramid coordination the degeneracy of the Cu 3$d$  orbitals is lifted in similar way as for an elongated octahedron. However, the fact that the pyramid apex is lying in the $z = 0$ plane implies that, to a first approximation, the hole is in the $d_{x^2-z^2}$ orbital. The $d_{x^2-z^2}$ orbital with only two lobes in the $ab$ plane while the two other lobes sticking out along the $c$ axis, is different from the $d_{x^2-y^2}$ orbital with all four lobes lying in the $ab$ plane, usually found in other cuprate superconductors. This should affect the strength of the magnetic couplings, which could be different along the $a$ and $b$ directions. It has to be noted, however, that the compressive distortion along the $c$ axis (with the Cu-O distances being equal along $a$ and $b$ but shorter along $c$) causes a further reduction of the point group symmetry of the Cu(2) ions from $D_{4h}$ to $D_{2h}$, with two consequences: the degeneracy of all 3$d$ states is removed, including that of the $d_{xy}$ and $d_{yz}$ orbitals, and the $d_{x^2-z^2}$ orbital is mixed up with the $d_{3y^2-r^2}$ orbital, adding some in-plane character to the hole wave function. With this in mind, we calculated the RIXS cross-sections\cite{ddMoretti} for a generic $\cos\alpha\ket{d_{x^2-z^2}}+\sin\alpha\ket{d_{3y^2-r^2}}$ ground state and fit the experimental RIXS spectra to the corresponding four excited states in order to extract the mixing parameter $\alpha$ and the crystal field splittings. Unfortunately, the problem is ill-defined because the $dd$ peak is very narrow and symmetric, and the decomposition is not unique. Therefore, at this stage, we can only state that the ground state has a mixed character of $d_{x^2-z^2}$ and $d_{3y^2-r^2}$ orbitals but we cannot determine the degree of admixture, for which the use of a single crystalline sample would be necessary. When single crystals of BCO will be available both XAS and RIXS will provide much more insightful information on the crystalline and electronic structure thanks to the polarization dependence of cross sections.
	
\begin{figure}[htbp!]
		\centering
		\includegraphics[width=1\columnwidth]{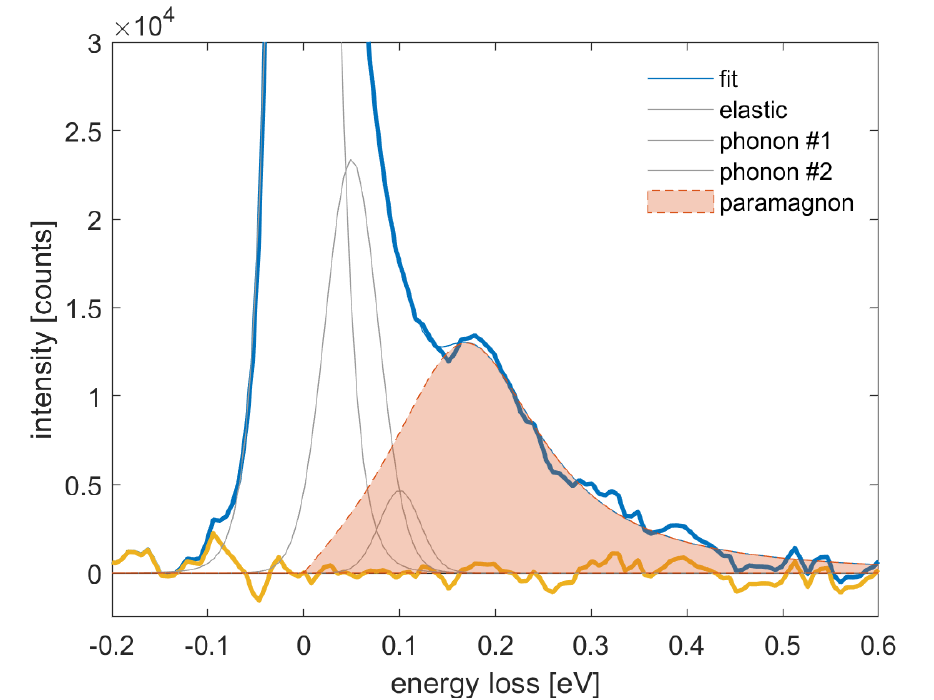}
		\caption{Close-up view of the RIXS spectrum of powder BCO measured at 932.7~eV incident photon energy (solid blue line) and its fit to an elastic line, two low-energy modes tentatively assigned to phonons (grey lines) and an asymmetric paramagnetic contribution (dashed red line with shaded area).}
		\label{fig_RIXS_magnetic}
	\end{figure}

	Finally, we discuss the low-energy excitations in BCO. The RIXS spectrum of Fig.~\ref{fig_RIXS_magnetic} is simultaneously fit to three symmetrical, energy resolution-limited curves (grey lines) and one asymmetrical curve (dashed red line); in agreement with existing literature on cuprates\cite{Rossi_EPC2019,Braicovich_2019arxiv}, the peaks at 60 and 110~meV are tentatively assigned to phonons and the peak at 180~meV to magnetic excitations (paramagnons)\cite{Braicovich_2010_magnonRIXS,MLTparamagnonsNatPhys}. Based on this interpretation, we will try in the following to give an estimate of the magnetic interactions in BCO.
	
	The proposed $ab$-plane crystal structure of BCO is shown in the inset of Fig.~\ref{fig_RIXS_powder_sim}(b), where Cu-O-Cu superexchange paths exist between nearest-neighbour Cu$^{2+}$ ions along three bonds only, and might be different along the $a$ and $b$ directions. The fourth magnetic bond is broken by the systematic absence of oxygen ions, thus preventing the system from forming the square magnetic lattice common to all layered cuprates. Magnetic excitations are in general highly dispersive, i.e., their energy depends on the magnitude and orientation of the transferred momentum, so precautions should be taken before drawing conclusions on magnetic interactions because, for a powder sample, the RIXS signal is averaged over the fraction of reciprocal space accessible by the available momentum transfer. Assuming isotropic (Heisenberg) antiferromagnetic couplings $J_a$ and $J_b \leq J_a$ between nearest-neighbour spins 1/2 along the $a$ and $b$ directions, respectively, we calculated the structure factor of magnetic excitations using linear spin wave theory, as implemented in SpinW\cite{SpinW}. The results are reported in Fig.~\ref{fig_RIXS_powder_sim} for $J_b=J_a=J$: the momentum transfer dependence of the magnetic structure factor for a powder sample is shown in panel (a) to be bounded at $1.5J$, as one should expect from simple considerations; a cut at 0.91~\AA$^{-1}$ momentum transfer, corresponding to performing RIXS at $2\theta=154^{\circ}$ scattering angle, as in our experiment, is reported in panel (b) and mostly consists of a single peak at an energy of 1.5$J$. When, by supposing an in-plane anisotropy of the superexchange interaction due to the lower density of oxygen ions along $b$,  the magnetic coupling along the $b$ direction is progressively decreased with respect to $J_a$, the upper boundary of the powder-averaged magnetic structure factor decreases to approximately $J_a$. The comparison with our RIXS data for BCO, then, suggests a superexchange coupling constant $120 \leq J_a \leq 180$~meV, in line with the magnitude of the antiferromagnetic superexchange interaction in most cuprates. We note that the BCO structure proposed here is very similar to that chosen by Wang et al.~to study the magnetic and electronic structure and superconductivity of this material with the $t-J$ model in the co-called brick-wall lattice \cite{Wang2019tj} of CuO$_{1.5}$ planes. 
	
	\section{Conclusions}
	\label{Conclusions}
	
	In our work, we probe the electronic structure of BCO by combining XAS and RIXS techniques. We argue that our measurements are only partially consistent with the crystal structure previously proposed for BCO and propose an alternative structure that is compatible with both previous pXRD measurements and recent theoretical calculations. The proposed structure features two inequivalent Cu sites and naturally justifies the double peak structure in the XAS profile. Moreover, the systematic vacancies of oxygens in the $z=0$ layer explains the absence of a high-energy crystal field excitation in the RIXS spectra, and leads to an estimated nearest neighbor superexchange intergrals in BCO in line with most common cuprates, i.e., of the order of 150~meV. Interestingly the in plane antiferromagnetic  square lattice is expected to characterize the superconducting planes of BCO as in the other cuprate superconductors, despite the loss of local $C_4$ symmetry in the magnetic interaction in BCO.
	
	From the structure here proposed, we predict that superconductivity occurs in the $z = 0$ layer and the ground state is an admixture of $d_{y^2-z^2}$ and $d_{3x^2-r^2}$ orbitals. Unfortunately, our present experimental data on powder BCO are insufficient to determine the degree of this mixture, but will possibly serve as a basis for future investigations on single crystal BCO samples. More in general, and beside the specific case of BCO, we demonstrate that XAS and RIXS can be usefully employed to set constraints on the plausible crystal structure of a newly synthesised oxide.

	\section{Acknowledgments}
	
	We acknowledge C.~Q.~Jin and W.~M.~Li for providing the sample, Z.~Hu for enlightening discussions and the Diamond Light Source for providing the beamtime on Beamline I21 under proposals NT23413. M.~Moretti Sala and G.~Ghiringhelli acknowledge support by project PRIN2017 ``Quantum-2D'' ID 2017Z8TS5B of the Ministry for University and Research (MIUR) of Italy.
	
	\bibliography{Bib_BCO}

\begin{thebibliography}{33}%
\makeatletter
\providecommand \@ifxundefined [1]{%
 \@ifx{#1\undefined}
}%
\providecommand \@ifnum [1]{%
 \ifnum #1\expandafter \@firstoftwo
 \else \expandafter \@secondoftwo
 \fi
}%
\providecommand \@ifx [1]{%
 \ifx #1\expandafter \@firstoftwo
 \else \expandafter \@secondoftwo
 \fi
}%
\providecommand \natexlab [1]{#1}%
\providecommand \enquote  [1]{``#1''}%
\providecommand \bibnamefont  [1]{#1}%
\providecommand \bibfnamefont [1]{#1}%
\providecommand \citenamefont [1]{#1}%
\providecommand \href@noop [0]{\@secondoftwo}%
\providecommand \href [0]{\begingroup \@sanitize@url \@href}%
\providecommand \@href[1]{\@@startlink{#1}\@@href}%
\providecommand \@@href[1]{\endgroup#1\@@endlink}%
\providecommand \@sanitize@url [0]{\catcode `\\12\catcode `\$12\catcode
  `\&12\catcode `\#12\catcode `\^12\catcode `\_12\catcode `\%12\relax}%
\providecommand \@@startlink[1]{}%
\providecommand \@@endlink[0]{}%
\providecommand \url  [0]{\begingroup\@sanitize@url \@url }%
\providecommand \@url [1]{\endgroup\@href {#1}{\urlprefix }}%
\providecommand \urlprefix  [0]{URL }%
\providecommand \Eprint [0]{\href }%
\providecommand \doibase [0]{http://dx.doi.org/}%
\providecommand \selectlanguage [0]{\@gobble}%
\providecommand \bibinfo  [0]{\@secondoftwo}%
\providecommand \bibfield  [0]{\@secondoftwo}%
\providecommand \translation [1]{[#1]}%
\providecommand \BibitemOpen [0]{}%
\providecommand \bibitemStop [0]{}%
\providecommand \bibitemNoStop [0]{.\EOS\space}%
\providecommand \EOS [0]{\spacefactor3000\relax}%
\providecommand \BibitemShut  [1]{\csname bibitem#1\endcsname}%
\let\auto@bib@innerbib\@empty
%</preamble>
\bibitem [{\citenamefont {Bednorz}\ and\ \citenamefont
  {Muller}(1986)}]{BednorzMuller1986}%
  \BibitemOpen
  \bibfield  {author} {\bibinfo {author} {\bibfnamefont {J.}~\bibnamefont
  {Bednorz}}\ and\ \bibinfo {author} {\bibfnamefont {K.}~\bibnamefont
  {Muller}},\ }\href {\doibase 10.1007/BF01303701} {\bibfield  {journal}
  {\bibinfo  {journal} {Z. Phys. B}\ }\textbf {\bibinfo {volume} {64}},\
  \bibinfo {pages} {189} (\bibinfo {year} {1986})}\BibitemShut {NoStop}%
\bibitem [{\citenamefont {Keimer}\ \emph {et~al.}(2014)\citenamefont {Keimer},
  \citenamefont {Kivelson}, \citenamefont {Norman}, \citenamefont {Uchida},\
  and\ \citenamefont {Zaanen}}]{KeimerKivelsonReviewHTS}%
  \BibitemOpen
  \bibfield  {author} {\bibinfo {author} {\bibfnamefont {B.}~\bibnamefont
  {Keimer}}, \bibinfo {author} {\bibfnamefont {S.~A.}\ \bibnamefont
  {Kivelson}}, \bibinfo {author} {\bibfnamefont {M.~R.}\ \bibnamefont
  {Norman}}, \bibinfo {author} {\bibfnamefont {S.}~\bibnamefont {Uchida}}, \
  and\ \bibinfo {author} {\bibfnamefont {J.}~\bibnamefont {Zaanen}},\ }\href
  {\doibase 10.1038/nature14165} {\bibfield  {journal} {\bibinfo  {journal}
  {Nature}\ }\textbf {\bibinfo {volume} {518}},\ \bibinfo {pages} {179}
  (\bibinfo {year} {2014})}\BibitemShut {NoStop}%
\bibitem [{\citenamefont {Li}\ \emph {et~al.}(2019{\natexlab{a}})\citenamefont
  {Li}, \citenamefont {Lee}, \citenamefont {Wang}, \citenamefont {Osada},
  \citenamefont {Crossley}, \citenamefont {Lee}, \citenamefont {Cui},
  \citenamefont {Hikita},\ and\ \citenamefont {Hwang}}]{Li2019_NickelateSC}%
  \BibitemOpen
  \bibfield  {author} {\bibinfo {author} {\bibfnamefont {D.}~\bibnamefont
  {Li}}, \bibinfo {author} {\bibfnamefont {K.}~\bibnamefont {Lee}}, \bibinfo
  {author} {\bibfnamefont {B.~Y.}\ \bibnamefont {Wang}}, \bibinfo {author}
  {\bibfnamefont {M.}~\bibnamefont {Osada}}, \bibinfo {author} {\bibfnamefont
  {S.}~\bibnamefont {Crossley}}, \bibinfo {author} {\bibfnamefont {H.~R.}\
  \bibnamefont {Lee}}, \bibinfo {author} {\bibfnamefont {Y.}~\bibnamefont
  {Cui}}, \bibinfo {author} {\bibfnamefont {Y.}~\bibnamefont {Hikita}}, \ and\
  \bibinfo {author} {\bibfnamefont {H.~Y.}\ \bibnamefont {Hwang}},\ }\href@noop
  {} {\bibfield  {journal} {\bibinfo  {journal} {Nature}\ }\textbf {\bibinfo
  {volume} {572}},\ \bibinfo {pages} {624} (\bibinfo {year}
  {2019}{\natexlab{a}})}\BibitemShut {NoStop}%
\bibitem [{\citenamefont {Wu}\ \emph {et~al.}(2019)\citenamefont {Wu},
  \citenamefont {Di~Sante}, \citenamefont {Schwemmer}, \citenamefont {Hanke},
  \citenamefont {Hwang}, \citenamefont {Raghu},\ and\ \citenamefont
  {Thomale}}]{Wu2019_arxiv}%
  \BibitemOpen
  \bibfield  {author} {\bibinfo {author} {\bibfnamefont {X.}~\bibnamefont
  {Wu}}, \bibinfo {author} {\bibfnamefont {D.}~\bibnamefont {Di~Sante}},
  \bibinfo {author} {\bibfnamefont {T.}~\bibnamefont {Schwemmer}}, \bibinfo
  {author} {\bibfnamefont {W.}~\bibnamefont {Hanke}}, \bibinfo {author}
  {\bibfnamefont {H.~Y.}\ \bibnamefont {Hwang}}, \bibinfo {author}
  {\bibfnamefont {S.}~\bibnamefont {Raghu}}, \ and\ \bibinfo {author}
  {\bibfnamefont {R.}~\bibnamefont {Thomale}},\ }\href@noop {} {\bibfield
  {journal} {\bibinfo  {journal} {arXiv preprint arXiv:1909.03015}\ } (\bibinfo
  {year} {2019})}\BibitemShut {NoStop}%
\bibitem [{\citenamefont {Sakakibara}\ \emph {et~al.}(2019)\citenamefont
  {Sakakibara}, \citenamefont {Usui}, \citenamefont {Suzuki}, \citenamefont
  {Kotani}, \citenamefont {Aoki},\ and\ \citenamefont
  {Kuroki}}]{Sakakibara2019_arxiv}%
  \BibitemOpen
  \bibfield  {author} {\bibinfo {author} {\bibfnamefont {H.}~\bibnamefont
  {Sakakibara}}, \bibinfo {author} {\bibfnamefont {H.}~\bibnamefont {Usui}},
  \bibinfo {author} {\bibfnamefont {K.}~\bibnamefont {Suzuki}}, \bibinfo
  {author} {\bibfnamefont {T.}~\bibnamefont {Kotani}}, \bibinfo {author}
  {\bibfnamefont {H.}~\bibnamefont {Aoki}}, \ and\ \bibinfo {author}
  {\bibfnamefont {K.}~\bibnamefont {Kuroki}},\ }\href@noop {} {\bibfield
  {journal} {\bibinfo  {journal} {arXiv preprint arXiv:1909.00060}\ } (\bibinfo
  {year} {2019})}\BibitemShut {NoStop}%
\bibitem [{\citenamefont {Botana}\ and\ \citenamefont
  {Norman}(2019)}]{Botana2019_arxiv}%
  \BibitemOpen
  \bibfield  {author} {\bibinfo {author} {\bibfnamefont {A.~S.}\ \bibnamefont
  {Botana}}\ and\ \bibinfo {author} {\bibfnamefont {M.~R.}\ \bibnamefont
  {Norman}},\ }\href@noop {} {\bibfield  {journal} {\bibinfo  {journal} {arXiv
  preprint arXiv:1908.10946}\ } (\bibinfo {year} {2019})}\BibitemShut {NoStop}%
\bibitem [{\citenamefont {Jiang}\ \emph {et~al.}(2019)\citenamefont {Jiang},
  \citenamefont {Berciu},\ and\ \citenamefont {Sawatzky}}]{Jiang2019_arxiv}%
  \BibitemOpen
  \bibfield  {author} {\bibinfo {author} {\bibfnamefont {M.}~\bibnamefont
  {Jiang}}, \bibinfo {author} {\bibfnamefont {M.}~\bibnamefont {Berciu}}, \
  and\ \bibinfo {author} {\bibfnamefont {G.~A.}\ \bibnamefont {Sawatzky}},\
  }\href@noop {} {\bibfield  {journal} {\bibinfo  {journal} {arXiv preprint
  arXiv:1909.02557}\ } (\bibinfo {year} {2019})}\BibitemShut {NoStop}%
\bibitem [{\citenamefont {Li}\ \emph {et~al.}(2019{\natexlab{b}})\citenamefont
  {Li}, \citenamefont {Zhao}, \citenamefont {Cao}, \citenamefont {Hu},
  \citenamefont {Huang}, \citenamefont {Wang}, \citenamefont {Liu},
  \citenamefont {Zhao}, \citenamefont {Zhang}, \citenamefont {Liu},
  \citenamefont {Yu}, \citenamefont {Long}, \citenamefont {Wu}, \citenamefont
  {Lin}, \citenamefont {Chen}, \citenamefont {Li}, \citenamefont {Gong},
  \citenamefont {Guguchia}, \citenamefont {Kim}, \citenamefont {Stewart},
  \citenamefont {Uemura}, \citenamefont {Uchida},\ and\ \citenamefont
  {Jin}}]{Li_2019}%
  \BibitemOpen
  \bibfield  {author} {\bibinfo {author} {\bibfnamefont {W.~M.}\ \bibnamefont
  {Li}}, \bibinfo {author} {\bibfnamefont {J.~F.}\ \bibnamefont {Zhao}},
  \bibinfo {author} {\bibfnamefont {L.~P.}\ \bibnamefont {Cao}}, \bibinfo
  {author} {\bibfnamefont {Z.}~\bibnamefont {Hu}}, \bibinfo {author}
  {\bibfnamefont {Q.~Z.}\ \bibnamefont {Huang}}, \bibinfo {author}
  {\bibfnamefont {X.~C.}\ \bibnamefont {Wang}}, \bibinfo {author}
  {\bibfnamefont {Y.}~\bibnamefont {Liu}}, \bibinfo {author} {\bibfnamefont
  {G.~Q.}\ \bibnamefont {Zhao}}, \bibinfo {author} {\bibfnamefont
  {J.}~\bibnamefont {Zhang}}, \bibinfo {author} {\bibfnamefont {Q.~Q.}\
  \bibnamefont {Liu}}, \bibinfo {author} {\bibfnamefont {R.~Z.}\ \bibnamefont
  {Yu}}, \bibinfo {author} {\bibfnamefont {Y.~W.}\ \bibnamefont {Long}},
  \bibinfo {author} {\bibfnamefont {H.}~\bibnamefont {Wu}}, \bibinfo {author}
  {\bibfnamefont {H.~J.}\ \bibnamefont {Lin}}, \bibinfo {author} {\bibfnamefont
  {C.~T.}\ \bibnamefont {Chen}}, \bibinfo {author} {\bibfnamefont
  {Z.}~\bibnamefont {Li}}, \bibinfo {author} {\bibfnamefont {Z.~Z.}\
  \bibnamefont {Gong}}, \bibinfo {author} {\bibfnamefont {Z.}~\bibnamefont
  {Guguchia}}, \bibinfo {author} {\bibfnamefont {J.~S.}\ \bibnamefont {Kim}},
  \bibinfo {author} {\bibfnamefont {G.~R.}\ \bibnamefont {Stewart}}, \bibinfo
  {author} {\bibfnamefont {Y.~J.}\ \bibnamefont {Uemura}}, \bibinfo {author}
  {\bibfnamefont {S.}~\bibnamefont {Uchida}}, \ and\ \bibinfo {author}
  {\bibfnamefont {C.~Q.}\ \bibnamefont {Jin}},\ }\href {\doibase
  10.1073/pnas.1900908116} {\bibfield  {journal} {\bibinfo  {journal}
  {Proceedings of the National Academy of Sciences}\ }\textbf {\bibinfo
  {volume} {116}},\ \bibinfo {pages} {12156} (\bibinfo {year}
  {2019}{\natexlab{b}})}\BibitemShut {NoStop}%
\bibitem [{\citenamefont {Chen}\ \emph {et~al.}(1992)\citenamefont {Chen},
  \citenamefont {Tjeng}, \citenamefont {Kwo}, \citenamefont {Kao},
  \citenamefont {Rudolf}, \citenamefont {Sette},\ and\ \citenamefont
  {Fleming}}]{Chen_1992}%
  \BibitemOpen
  \bibfield  {author} {\bibinfo {author} {\bibfnamefont {C.~T.}\ \bibnamefont
  {Chen}}, \bibinfo {author} {\bibfnamefont {L.~H.}\ \bibnamefont {Tjeng}},
  \bibinfo {author} {\bibfnamefont {J.}~\bibnamefont {Kwo}}, \bibinfo {author}
  {\bibfnamefont {H.~L.}\ \bibnamefont {Kao}}, \bibinfo {author} {\bibfnamefont
  {P.}~\bibnamefont {Rudolf}}, \bibinfo {author} {\bibfnamefont
  {F.}~\bibnamefont {Sette}}, \ and\ \bibinfo {author} {\bibfnamefont {R.~M.}\
  \bibnamefont {Fleming}},\ }\href {\doibase 10.1103/PhysRevLett.68.2543}
  {\bibfield  {journal} {\bibinfo  {journal} {Phys. Rev. Lett.}\ }\textbf
  {\bibinfo {volume} {68}},\ \bibinfo {pages} {2543} (\bibinfo {year}
  {1992})}\BibitemShut {NoStop}%
\bibitem [{\citenamefont {Peng}\ \emph {et~al.}(2017)\citenamefont {Peng},
  \citenamefont {Dellea}, \citenamefont {Minola}, \citenamefont {Conni},
  \citenamefont {Amorese}, \citenamefont {Castro}, \citenamefont {Luca},
  \citenamefont {Kummer}, \citenamefont {Salluzzo}, \citenamefont {Sun},
  \citenamefont {Zhou}, \citenamefont {Balestrino}, \citenamefont {Tacon},
  \citenamefont {Keimer}, \citenamefont {Braicovich}, \citenamefont {Brookes},\
  and\ \citenamefont {Ghiringhelli}}]{Peng_NatPhys}%
  \BibitemOpen
  \bibfield  {author} {\bibinfo {author} {\bibfnamefont {Y.~Y.}\ \bibnamefont
  {Peng}}, \bibinfo {author} {\bibfnamefont {G.}~\bibnamefont {Dellea}},
  \bibinfo {author} {\bibfnamefont {M.}~\bibnamefont {Minola}}, \bibinfo
  {author} {\bibfnamefont {M.}~\bibnamefont {Conni}}, \bibinfo {author}
  {\bibfnamefont {A.}~\bibnamefont {Amorese}}, \bibinfo {author} {\bibfnamefont
  {D.~D.}\ \bibnamefont {Castro}}, \bibinfo {author} {\bibfnamefont {G.~M.~D.}\
  \bibnamefont {Luca}}, \bibinfo {author} {\bibfnamefont {K.}~\bibnamefont
  {Kummer}}, \bibinfo {author} {\bibfnamefont {M.}~\bibnamefont {Salluzzo}},
  \bibinfo {author} {\bibfnamefont {X.}~\bibnamefont {Sun}}, \bibinfo {author}
  {\bibfnamefont {X.~J.}\ \bibnamefont {Zhou}}, \bibinfo {author}
  {\bibfnamefont {G.}~\bibnamefont {Balestrino}}, \bibinfo {author}
  {\bibfnamefont {M.~L.}\ \bibnamefont {Tacon}}, \bibinfo {author}
  {\bibfnamefont {B.}~\bibnamefont {Keimer}}, \bibinfo {author} {\bibfnamefont
  {L.}~\bibnamefont {Braicovich}}, \bibinfo {author} {\bibfnamefont {N.~B.}\
  \bibnamefont {Brookes}}, \ and\ \bibinfo {author} {\bibfnamefont
  {G.}~\bibnamefont {Ghiringhelli}},\ }\href {\doibase 10.1038/NPHYS4248}
  {\bibfield  {journal} {\bibinfo  {journal} {Nat. Phys.}\ }\textbf {\bibinfo
  {volume} {13}},\ \bibinfo {pages} {1201} (\bibinfo {year}
  {2017})}\BibitemShut {NoStop}%
\bibitem [{\citenamefont {Maier}\ \emph {et~al.}(2019)\citenamefont {Maier},
  \citenamefont {Berlijn},\ and\ \citenamefont {Scalapino}}]{Maier_PRB2019}%
  \BibitemOpen
  \bibfield  {author} {\bibinfo {author} {\bibfnamefont {T.}~\bibnamefont
  {Maier}}, \bibinfo {author} {\bibfnamefont {T.}~\bibnamefont {Berlijn}}, \
  and\ \bibinfo {author} {\bibfnamefont {D.~J.}\ \bibnamefont {Scalapino}},\
  }\href {\doibase 10.1103/PhysRevB.99.224515} {\bibfield  {journal} {\bibinfo
  {journal} {Phys. Rev. B}\ }\textbf {\bibinfo {volume} {99}},\ \bibinfo
  {pages} {224515} (\bibinfo {year} {2019})}\BibitemShut {NoStop}%
\bibitem [{\citenamefont {{Moretti Sala}}\ \emph {et~al.}(2011)\citenamefont
  {{Moretti Sala}}, \citenamefont {Bisogni}, \citenamefont {Aruta},
  \citenamefont {Balestrino}, \citenamefont {Berger}, \citenamefont {Brookes},
  \citenamefont {{De Luca}}, \citenamefont {{Di Castro}}, \citenamefont
  {Grioni}, \citenamefont {Guarise}, \citenamefont {Medaglia}, \citenamefont
  {{Miletto Granozio}}, \citenamefont {Minola}, \citenamefont {Perna},
  \citenamefont {Radovic}, \citenamefont {Salluzzo}, \citenamefont {Schmitt},
  \citenamefont {Zhou}, \citenamefont {Braicovich},\ and\ \citenamefont
  {Ghiringhelli}}]{ddMoretti}%
  \BibitemOpen
  \bibfield  {author} {\bibinfo {author} {\bibfnamefont {M.}~\bibnamefont
  {{Moretti Sala}}}, \bibinfo {author} {\bibfnamefont {V.}~\bibnamefont
  {Bisogni}}, \bibinfo {author} {\bibfnamefont {C.}~\bibnamefont {Aruta}},
  \bibinfo {author} {\bibfnamefont {G.}~\bibnamefont {Balestrino}}, \bibinfo
  {author} {\bibfnamefont {H.}~\bibnamefont {Berger}}, \bibinfo {author}
  {\bibfnamefont {N.~B.}\ \bibnamefont {Brookes}}, \bibinfo {author}
  {\bibfnamefont {G.~M.}\ \bibnamefont {{De Luca}}}, \bibinfo {author}
  {\bibfnamefont {D.}~\bibnamefont {{Di Castro}}}, \bibinfo {author}
  {\bibfnamefont {M.}~\bibnamefont {Grioni}}, \bibinfo {author} {\bibfnamefont
  {M.}~\bibnamefont {Guarise}}, \bibinfo {author} {\bibfnamefont {P.~G.}\
  \bibnamefont {Medaglia}}, \bibinfo {author} {\bibfnamefont {F.}~\bibnamefont
  {{Miletto Granozio}}}, \bibinfo {author} {\bibfnamefont {M.}~\bibnamefont
  {Minola}}, \bibinfo {author} {\bibfnamefont {P.}~\bibnamefont {Perna}},
  \bibinfo {author} {\bibfnamefont {M.}~\bibnamefont {Radovic}}, \bibinfo
  {author} {\bibfnamefont {M.}~\bibnamefont {Salluzzo}}, \bibinfo {author}
  {\bibfnamefont {T.}~\bibnamefont {Schmitt}}, \bibinfo {author} {\bibfnamefont
  {K.~J.}\ \bibnamefont {Zhou}}, \bibinfo {author} {\bibfnamefont
  {L.}~\bibnamefont {Braicovich}}, \ and\ \bibinfo {author} {\bibfnamefont
  {G.}~\bibnamefont {Ghiringhelli}},\ }\href {\doibase
  10.1088/1367-2630/13/4/043026} {\bibfield  {journal} {\bibinfo  {journal}
  {New J. Phys.}\ }\textbf {\bibinfo {volume} {13}},\ \bibinfo {pages} {043026}
  (\bibinfo {year} {2011})}\BibitemShut {NoStop}%
\bibitem [{\citenamefont {Ament}\ \emph {et~al.}(2011)\citenamefont {Ament},
  \citenamefont {van Veenendaal}, \citenamefont {Devereaux}, \citenamefont
  {Hill},\ and\ \citenamefont {van~den Brink}}]{Ament_review}%
  \BibitemOpen
  \bibfield  {author} {\bibinfo {author} {\bibfnamefont {L.~J.~P.}\
  \bibnamefont {Ament}}, \bibinfo {author} {\bibfnamefont {M.}~\bibnamefont
  {van Veenendaal}}, \bibinfo {author} {\bibfnamefont {T.~P.}\ \bibnamefont
  {Devereaux}}, \bibinfo {author} {\bibfnamefont {J.~P.}\ \bibnamefont {Hill}},
  \ and\ \bibinfo {author} {\bibfnamefont {J.}~\bibnamefont {van~den Brink}},\
  }\href {\doibase 10.1103/RevModPhys.83.705} {\bibfield  {journal} {\bibinfo
  {journal} {Rev. Mod. Phys.}\ }\textbf {\bibinfo {volume} {83}},\ \bibinfo
  {pages} {705} (\bibinfo {year} {2011})}\BibitemShut {NoStop}%
\bibitem [{\citenamefont {Minola}\ \emph {et~al.}(2015)\citenamefont {Minola},
  \citenamefont {Dellea}, \citenamefont {Gretarsson}, \citenamefont {Peng},
  \citenamefont {Lu}, \citenamefont {Porras}, \citenamefont {Loew},
  \citenamefont {Yakhou}, \citenamefont {Brookes}, \citenamefont {Huang},
  \citenamefont {Pelliciari}, \citenamefont {Schmitt}, \citenamefont
  {Ghiringhelli}, \citenamefont {Keimer}, \citenamefont {Braicovich},\ and\
  \citenamefont {Le~Tacon}}]{Minola_PRL2015}%
  \BibitemOpen
  \bibfield  {author} {\bibinfo {author} {\bibfnamefont {M.}~\bibnamefont
  {Minola}}, \bibinfo {author} {\bibfnamefont {G.}~\bibnamefont {Dellea}},
  \bibinfo {author} {\bibfnamefont {H.}~\bibnamefont {Gretarsson}}, \bibinfo
  {author} {\bibfnamefont {Y.~Y.}\ \bibnamefont {Peng}}, \bibinfo {author}
  {\bibfnamefont {Y.}~\bibnamefont {Lu}}, \bibinfo {author} {\bibfnamefont
  {J.}~\bibnamefont {Porras}}, \bibinfo {author} {\bibfnamefont
  {T.}~\bibnamefont {Loew}}, \bibinfo {author} {\bibfnamefont {F.}~\bibnamefont
  {Yakhou}}, \bibinfo {author} {\bibfnamefont {N.~B.}\ \bibnamefont {Brookes}},
  \bibinfo {author} {\bibfnamefont {Y.~B.}\ \bibnamefont {Huang}}, \bibinfo
  {author} {\bibfnamefont {J.}~\bibnamefont {Pelliciari}}, \bibinfo {author}
  {\bibfnamefont {T.}~\bibnamefont {Schmitt}}, \bibinfo {author} {\bibfnamefont
  {G.}~\bibnamefont {Ghiringhelli}}, \bibinfo {author} {\bibfnamefont
  {B.}~\bibnamefont {Keimer}}, \bibinfo {author} {\bibfnamefont
  {L.}~\bibnamefont {Braicovich}}, \ and\ \bibinfo {author} {\bibfnamefont
  {M.}~\bibnamefont {Le~Tacon}},\ }\href {\doibase
  10.1103/PhysRevLett.114.217003} {\bibfield  {journal} {\bibinfo  {journal}
  {Phys. Rev. Lett.}\ }\textbf {\bibinfo {volume} {114}},\ \bibinfo {pages}
  {217003} (\bibinfo {year} {2015})}\BibitemShut {NoStop}%
\bibitem [{\citenamefont {Hawthorn}\ \emph {et~al.}(2011)\citenamefont
  {Hawthorn}, \citenamefont {Shen}, \citenamefont {Geck}, \citenamefont
  {Peets}, \citenamefont {Wadati}, \citenamefont {Okamoto}, \citenamefont
  {Huang}, \citenamefont {Huang}, \citenamefont {Lin}, \citenamefont
  {Denlinger}, \citenamefont {Liang}, \citenamefont {Bonn}, \citenamefont
  {Hardy},\ and\ \citenamefont {Sawatzky}}]{Hawthorn_2011}%
  \BibitemOpen
  \bibfield  {author} {\bibinfo {author} {\bibfnamefont {D.~G.}\ \bibnamefont
  {Hawthorn}}, \bibinfo {author} {\bibfnamefont {K.~M.}\ \bibnamefont {Shen}},
  \bibinfo {author} {\bibfnamefont {J.}~\bibnamefont {Geck}}, \bibinfo {author}
  {\bibfnamefont {D.~C.}\ \bibnamefont {Peets}}, \bibinfo {author}
  {\bibfnamefont {H.}~\bibnamefont {Wadati}}, \bibinfo {author} {\bibfnamefont
  {J.}~\bibnamefont {Okamoto}}, \bibinfo {author} {\bibfnamefont {S.-W.}\
  \bibnamefont {Huang}}, \bibinfo {author} {\bibfnamefont {D.~J.}\ \bibnamefont
  {Huang}}, \bibinfo {author} {\bibfnamefont {H.-J.}\ \bibnamefont {Lin}},
  \bibinfo {author} {\bibfnamefont {J.~D.}\ \bibnamefont {Denlinger}}, \bibinfo
  {author} {\bibfnamefont {R.}~\bibnamefont {Liang}}, \bibinfo {author}
  {\bibfnamefont {D.~A.}\ \bibnamefont {Bonn}}, \bibinfo {author}
  {\bibfnamefont {W.~N.}\ \bibnamefont {Hardy}}, \ and\ \bibinfo {author}
  {\bibfnamefont {G.~A.}\ \bibnamefont {Sawatzky}},\ }\href {\doibase
  10.1103/PhysRevB.84.075125} {\bibfield  {journal} {\bibinfo  {journal} {Phys.
  Rev. B}\ }\textbf {\bibinfo {volume} {84}},\ \bibinfo {pages} {075125}
  (\bibinfo {year} {2011})}\BibitemShut {NoStop}%
\bibitem [{\citenamefont {Dean}\ \emph {et~al.}(2013)\citenamefont {Dean},
  \citenamefont {Dellea}, \citenamefont {Springell}, \citenamefont
  {Yakhou-Harris}, \citenamefont {Kummer}, \citenamefont {Brookes},
  \citenamefont {Liu}, \citenamefont {Sun}, \citenamefont {Strle},
  \citenamefont {Schmitt} \emph {et~al.}}]{Dean_2013}%
  \BibitemOpen
  \bibfield  {author} {\bibinfo {author} {\bibfnamefont {M.}~\bibnamefont
  {Dean}}, \bibinfo {author} {\bibfnamefont {G.}~\bibnamefont {Dellea}},
  \bibinfo {author} {\bibfnamefont {R.}~\bibnamefont {Springell}}, \bibinfo
  {author} {\bibfnamefont {F.}~\bibnamefont {Yakhou-Harris}}, \bibinfo {author}
  {\bibfnamefont {K.}~\bibnamefont {Kummer}}, \bibinfo {author} {\bibfnamefont
  {N.}~\bibnamefont {Brookes}}, \bibinfo {author} {\bibfnamefont
  {X.}~\bibnamefont {Liu}}, \bibinfo {author} {\bibfnamefont {Y.}~\bibnamefont
  {Sun}}, \bibinfo {author} {\bibfnamefont {J.}~\bibnamefont {Strle}}, \bibinfo
  {author} {\bibfnamefont {T.}~\bibnamefont {Schmitt}},  \emph {et~al.},\
  }\href {\doibase doi:10.1038/nmat3723} {\bibfield  {journal} {\bibinfo
  {journal} {Nat. Mat.}\ }\textbf {\bibinfo {volume} {12}},\ \bibinfo {pages}
  {1019} (\bibinfo {year} {2013})}\BibitemShut {NoStop}%
\bibitem [{\citenamefont {Kang}\ \emph {et~al.}(2019)\citenamefont {Kang},
  \citenamefont {Pelliciari}, \citenamefont {Krockenberger}, \citenamefont
  {Li}, \citenamefont {McNally}, \citenamefont {Paris}, \citenamefont {Liang},
  \citenamefont {Hardy}, \citenamefont {Bonn}, \citenamefont {Yamamoto},
  \citenamefont {Schmitt},\ and\ \citenamefont {Comin}}]{Kang_2019}%
  \BibitemOpen
  \bibfield  {author} {\bibinfo {author} {\bibfnamefont {M.}~\bibnamefont
  {Kang}}, \bibinfo {author} {\bibfnamefont {J.}~\bibnamefont {Pelliciari}},
  \bibinfo {author} {\bibfnamefont {Y.}~\bibnamefont {Krockenberger}}, \bibinfo
  {author} {\bibfnamefont {J.}~\bibnamefont {Li}}, \bibinfo {author}
  {\bibfnamefont {D.~E.}\ \bibnamefont {McNally}}, \bibinfo {author}
  {\bibfnamefont {E.}~\bibnamefont {Paris}}, \bibinfo {author} {\bibfnamefont
  {R.}~\bibnamefont {Liang}}, \bibinfo {author} {\bibfnamefont {W.~N.}\
  \bibnamefont {Hardy}}, \bibinfo {author} {\bibfnamefont {D.~A.}\ \bibnamefont
  {Bonn}}, \bibinfo {author} {\bibfnamefont {H.}~\bibnamefont {Yamamoto}},
  \bibinfo {author} {\bibfnamefont {T.}~\bibnamefont {Schmitt}}, \ and\
  \bibinfo {author} {\bibfnamefont {R.}~\bibnamefont {Comin}},\ }\href
  {\doibase 10.1103/PhysRevB.99.045105} {\bibfield  {journal} {\bibinfo
  {journal} {Phys. Rev. B}\ }\textbf {\bibinfo {volume} {99}},\ \bibinfo
  {pages} {045105} (\bibinfo {year} {2019})}\BibitemShut {NoStop}%
\bibitem [{\citenamefont {Fumagalli}\ \emph {et~al.}(2019)\citenamefont
  {Fumagalli}, \citenamefont {Braicovich}, \citenamefont {Minola},
  \citenamefont {Peng}, \citenamefont {Kummer}, \citenamefont {Betto},
  \citenamefont {Rossi}, \citenamefont {Lefran\ifmmode~\mbox{\c{c}}\else
  \c{c}\fi{}ois}, \citenamefont {Morawe}, \citenamefont {Salluzzo},
  \citenamefont {Suzuki}, \citenamefont {Yakhou}, \citenamefont {Le~Tacon},
  \citenamefont {Keimer}, \citenamefont {Brookes}, \citenamefont {Sala},\ and\
  \citenamefont {Ghiringhelli}}]{Fumagalli_2019}%
  \BibitemOpen
  \bibfield  {author} {\bibinfo {author} {\bibfnamefont {R.}~\bibnamefont
  {Fumagalli}}, \bibinfo {author} {\bibfnamefont {L.}~\bibnamefont
  {Braicovich}}, \bibinfo {author} {\bibfnamefont {M.}~\bibnamefont {Minola}},
  \bibinfo {author} {\bibfnamefont {Y.~Y.}\ \bibnamefont {Peng}}, \bibinfo
  {author} {\bibfnamefont {K.}~\bibnamefont {Kummer}}, \bibinfo {author}
  {\bibfnamefont {D.}~\bibnamefont {Betto}}, \bibinfo {author} {\bibfnamefont
  {M.}~\bibnamefont {Rossi}}, \bibinfo {author} {\bibfnamefont
  {E.}~\bibnamefont {Lefran\ifmmode~\mbox{\c{c}}\else \c{c}\fi{}ois}}, \bibinfo
  {author} {\bibfnamefont {C.}~\bibnamefont {Morawe}}, \bibinfo {author}
  {\bibfnamefont {M.}~\bibnamefont {Salluzzo}}, \bibinfo {author}
  {\bibfnamefont {H.}~\bibnamefont {Suzuki}}, \bibinfo {author} {\bibfnamefont
  {F.}~\bibnamefont {Yakhou}}, \bibinfo {author} {\bibfnamefont
  {M.}~\bibnamefont {Le~Tacon}}, \bibinfo {author} {\bibfnamefont
  {B.}~\bibnamefont {Keimer}}, \bibinfo {author} {\bibfnamefont {N.~B.}\
  \bibnamefont {Brookes}}, \bibinfo {author} {\bibfnamefont {M.~M.}\
  \bibnamefont {Sala}}, \ and\ \bibinfo {author} {\bibfnamefont
  {G.}~\bibnamefont {Ghiringhelli}},\ }\href {\doibase
  10.1103/PhysRevB.99.134517} {\bibfield  {journal} {\bibinfo  {journal} {Phys.
  Rev. B}\ }\textbf {\bibinfo {volume} {99}},\ \bibinfo {pages} {134517}
  (\bibinfo {year} {2019})}\BibitemShut {NoStop}%
\bibitem [{\citenamefont {Manske}(2004)}]{Manske_2004}%
  \BibitemOpen
  \bibfield  {author} {\bibinfo {author} {\bibfnamefont {D.}~\bibnamefont
  {Manske}},\ }\href@noop {} {\emph {\bibinfo {title} {Theory of unconventional
  superconductors: cooper-pairing mediated by spin excitations}}},\ Vol.\
  \bibinfo {volume} {202}\ (\bibinfo  {publisher} {Springer Science \& Business
  Media},\ \bibinfo {year} {2004})\BibitemShut {NoStop}%
\bibitem [{\citenamefont {Siegrist}\ \emph {et~al.}(1988)\citenamefont
  {Siegrist}, \citenamefont {Zahurak}, \citenamefont {Murphy},\ and\
  \citenamefont {Roth}}]{Siegrist_1988}%
  \BibitemOpen
  \bibfield  {author} {\bibinfo {author} {\bibfnamefont {T.}~\bibnamefont
  {Siegrist}}, \bibinfo {author} {\bibfnamefont {S.~M.}\ \bibnamefont
  {Zahurak}}, \bibinfo {author} {\bibfnamefont {D.~W.}\ \bibnamefont {Murphy}},
  \ and\ \bibinfo {author} {\bibfnamefont {R.~S.}\ \bibnamefont {Roth}},\
  }\href {\doibase 10.1038/334231a0} {\bibfield  {journal} {\bibinfo  {journal}
  {Nature}\ }\textbf {\bibinfo {volume} {334}},\ \bibinfo {pages} {231}
  (\bibinfo {year} {1988})}\BibitemShut {NoStop}%
\bibitem [{\citenamefont {Hozoi}\ \emph {et~al.}(2011)\citenamefont {Hozoi},
  \citenamefont {Siurakshina}, \citenamefont {Fulde},\ and\ \citenamefont {Van
  Den~Brink}}]{Hozoi_2011}%
  \BibitemOpen
  \bibfield  {author} {\bibinfo {author} {\bibfnamefont {L.}~\bibnamefont
  {Hozoi}}, \bibinfo {author} {\bibfnamefont {L.}~\bibnamefont {Siurakshina}},
  \bibinfo {author} {\bibfnamefont {P.}~\bibnamefont {Fulde}}, \ and\ \bibinfo
  {author} {\bibfnamefont {J.}~\bibnamefont {Van Den~Brink}},\ }\href {\doibase
  10.1038/srep00065} {\bibfield  {journal} {\bibinfo  {journal} {Scientific
  reports}\ }\textbf {\bibinfo {volume} {1}},\ \bibinfo {pages} {65} (\bibinfo
  {year} {2011})}\BibitemShut {NoStop}%
\bibitem [{\citenamefont {Bersuker}(2010)}]{Bersuker}%
  \BibitemOpen
  \bibfield  {author} {\bibinfo {author} {\bibfnamefont {I.~B.}\ \bibnamefont
  {Bersuker}},\ }\href {\doibase 10.1002/9780470573051} {\emph {\bibinfo
  {title} {{E}lectronic structure and properties of transition metal
  compounds}}}\ (\bibinfo  {publisher} {John Wiley \& Sons},\ \bibinfo {year}
  {2010})\BibitemShut {NoStop}%
\bibitem [{\citenamefont {Li}\ \emph {et~al.}(2019{\natexlab{c}})\citenamefont
  {Li}, \citenamefont {Du}, \citenamefont {Weng},\ and\ \citenamefont
  {Liu}}]{Li_2019arxiv}%
  \BibitemOpen
  \bibfield  {author} {\bibinfo {author} {\bibfnamefont {Y.}~\bibnamefont
  {Li}}, \bibinfo {author} {\bibfnamefont {S.}~\bibnamefont {Du}}, \bibinfo
  {author} {\bibfnamefont {Z.-Y.}\ \bibnamefont {Weng}}, \ and\ \bibinfo
  {author} {\bibfnamefont {Z.}~\bibnamefont {Liu}},\ }\href@noop {} {\bibfield
  {journal} {\bibinfo  {journal} {arXiv preprint arXiv:1909.08304}\ } (\bibinfo
  {year} {2019}{\natexlab{c}})}\BibitemShut {NoStop}%
\bibitem [{\citenamefont {Le}\ \emph {et~al.}(2019)\citenamefont {Le},
  \citenamefont {Jiang}, \citenamefont {Li}, \citenamefont {Qin}, \citenamefont
  {Wang}, \citenamefont {Zhang},\ and\ \citenamefont {Hu}}]{Le_2019_arxiv}%
  \BibitemOpen
  \bibfield  {author} {\bibinfo {author} {\bibfnamefont {C.}~\bibnamefont
  {Le}}, \bibinfo {author} {\bibfnamefont {K.}~\bibnamefont {Jiang}}, \bibinfo
  {author} {\bibfnamefont {Y.}~\bibnamefont {Li}}, \bibinfo {author}
  {\bibfnamefont {S.}~\bibnamefont {Qin}}, \bibinfo {author} {\bibfnamefont
  {Z.}~\bibnamefont {Wang}}, \bibinfo {author} {\bibfnamefont {F.}~\bibnamefont
  {Zhang}}, \ and\ \bibinfo {author} {\bibfnamefont {J.}~\bibnamefont {Hu}},\
  }\href@noop {} {\bibfield  {journal} {\bibinfo  {journal} {arXiv preprint
  arXiv:1909.12620}\ } (\bibinfo {year} {2019})}\BibitemShut {NoStop}%
\bibitem [{\citenamefont {Liu}\ \emph {et~al.}(2019)\citenamefont {Liu},
  \citenamefont {Lu},\ and\ \citenamefont {Xiang}}]{Liu_PhysRevMaterials2019}%
  \BibitemOpen
  \bibfield  {author} {\bibinfo {author} {\bibfnamefont {K.}~\bibnamefont
  {Liu}}, \bibinfo {author} {\bibfnamefont {Z.-Y.}\ \bibnamefont {Lu}}, \ and\
  \bibinfo {author} {\bibfnamefont {T.}~\bibnamefont {Xiang}},\ }\href
  {\doibase 10.1103/PhysRevMaterials.3.044802} {\bibfield  {journal} {\bibinfo
  {journal} {Phys. Rev. Materials}\ }\textbf {\bibinfo {volume} {3}},\ \bibinfo
  {pages} {044802} (\bibinfo {year} {2019})}\BibitemShut {NoStop}%
\bibitem [{\citenamefont {Merz}\ \emph {et~al.}(1998)\citenamefont {Merz},
  \citenamefont {N\"ucker}, \citenamefont {Schweiss}, \citenamefont
  {Schuppler}, \citenamefont {Chen}, \citenamefont {Chakarian}, \citenamefont
  {Freeland}, \citenamefont {Idzerda}, \citenamefont {Kl\"aser}, \citenamefont
  {M\"uller-Vogt},\ and\ \citenamefont {Wolf}}]{Merz1998}%
  \BibitemOpen
  \bibfield  {author} {\bibinfo {author} {\bibfnamefont {M.}~\bibnamefont
  {Merz}}, \bibinfo {author} {\bibfnamefont {N.}~\bibnamefont {N\"ucker}},
  \bibinfo {author} {\bibfnamefont {P.}~\bibnamefont {Schweiss}}, \bibinfo
  {author} {\bibfnamefont {S.}~\bibnamefont {Schuppler}}, \bibinfo {author}
  {\bibfnamefont {C.~T.}\ \bibnamefont {Chen}}, \bibinfo {author}
  {\bibfnamefont {V.}~\bibnamefont {Chakarian}}, \bibinfo {author}
  {\bibfnamefont {J.}~\bibnamefont {Freeland}}, \bibinfo {author}
  {\bibfnamefont {Y.~U.}\ \bibnamefont {Idzerda}}, \bibinfo {author}
  {\bibfnamefont {M.}~\bibnamefont {Kl\"aser}}, \bibinfo {author}
  {\bibfnamefont {G.}~\bibnamefont {M\"uller-Vogt}}, \ and\ \bibinfo {author}
  {\bibfnamefont {T.}~\bibnamefont {Wolf}},\ }\href {\doibase
  10.1103/PhysRevLett.80.5192} {\bibfield  {journal} {\bibinfo  {journal}
  {Phys. Rev. Lett.}\ }\textbf {\bibinfo {volume} {80}},\ \bibinfo {pages}
  {5192} (\bibinfo {year} {1998})}\BibitemShut {NoStop}%
\bibitem [{\citenamefont {Hu}\ \emph {et~al.}(2002)\citenamefont {Hu},
  \citenamefont {Drechsler}, \citenamefont {M{\'a}lek}, \citenamefont {Rosner},
  \citenamefont {Neudert}, \citenamefont {Knupfer}, \citenamefont {Golden},
  \citenamefont {Fink}, \citenamefont {Karpinski}, \citenamefont {Kaindl},
  \citenamefont {Hellwig},\ and\ \citenamefont {Jung}}]{Hu2002}%
  \BibitemOpen
  \bibfield  {author} {\bibinfo {author} {\bibfnamefont {Z.}~\bibnamefont
  {Hu}}, \bibinfo {author} {\bibfnamefont {S.-L.}\ \bibnamefont {Drechsler}},
  \bibinfo {author} {\bibfnamefont {J.}~\bibnamefont {M{\'a}lek}}, \bibinfo
  {author} {\bibfnamefont {H.}~\bibnamefont {Rosner}}, \bibinfo {author}
  {\bibfnamefont {R.}~\bibnamefont {Neudert}}, \bibinfo {author} {\bibfnamefont
  {M.}~\bibnamefont {Knupfer}}, \bibinfo {author} {\bibfnamefont {M.~S.}\
  \bibnamefont {Golden}}, \bibinfo {author} {\bibfnamefont {J.}~\bibnamefont
  {Fink}}, \bibinfo {author} {\bibfnamefont {J.}~\bibnamefont {Karpinski}},
  \bibinfo {author} {\bibfnamefont {G.}~\bibnamefont {Kaindl}}, \bibinfo
  {author} {\bibfnamefont {C.}~\bibnamefont {Hellwig}}, \ and\ \bibinfo
  {author} {\bibfnamefont {C.}~\bibnamefont {Jung}},\ }\href {\doibase
  10.1209/epl/i2002-00168-7} {\bibfield  {journal} {\bibinfo  {journal}
  {Europhysics Letters}\ }\textbf {\bibinfo {volume} {59}},\ \bibinfo {pages}
  {135} (\bibinfo {year} {2002})}\BibitemShut {NoStop}%
\bibitem [{\citenamefont {Rossi}\ \emph {et~al.}(2019)\citenamefont {Rossi},
  \citenamefont {Arpaia}, \citenamefont {Fumagalli}, \citenamefont
  {Moretti~Sala}, \citenamefont {Betto}, \citenamefont {Kummer}, \citenamefont
  {De~Luca}, \citenamefont {van~den Brink}, \citenamefont {Salluzzo},
  \citenamefont {Brookes}, \citenamefont {Braicovich},\ and\ \citenamefont
  {Ghiringhelli}}]{Rossi_EPC2019}%
  \BibitemOpen
  \bibfield  {author} {\bibinfo {author} {\bibfnamefont {M.}~\bibnamefont
  {Rossi}}, \bibinfo {author} {\bibfnamefont {R.}~\bibnamefont {Arpaia}},
  \bibinfo {author} {\bibfnamefont {R.}~\bibnamefont {Fumagalli}}, \bibinfo
  {author} {\bibfnamefont {M.}~\bibnamefont {Moretti~Sala}}, \bibinfo {author}
  {\bibfnamefont {D.}~\bibnamefont {Betto}}, \bibinfo {author} {\bibfnamefont
  {K.}~\bibnamefont {Kummer}}, \bibinfo {author} {\bibfnamefont {G.~M.}\
  \bibnamefont {De~Luca}}, \bibinfo {author} {\bibfnamefont {J.}~\bibnamefont
  {van~den Brink}}, \bibinfo {author} {\bibfnamefont {M.}~\bibnamefont
  {Salluzzo}}, \bibinfo {author} {\bibfnamefont {N.~B.}\ \bibnamefont
  {Brookes}}, \bibinfo {author} {\bibfnamefont {L.}~\bibnamefont {Braicovich}},
  \ and\ \bibinfo {author} {\bibfnamefont {G.}~\bibnamefont {Ghiringhelli}},\
  }\href {\doibase 10.1103/PhysRevLett.123.027001} {\bibfield  {journal}
  {\bibinfo  {journal} {Physical Review Letters}\ }\textbf {\bibinfo {volume}
  {123}},\ \bibinfo {pages} {027001} (\bibinfo {year} {2019})}\BibitemShut
  {NoStop}%
\bibitem [{\citenamefont {Braicovich}\ \emph {et~al.}(2019)\citenamefont
  {Braicovich}, \citenamefont {Rossi}, \citenamefont {Fumagalli}, \citenamefont
  {Peng}, \citenamefont {Wang}, \citenamefont {Arpaia}, \citenamefont {Betto},
  \citenamefont {De~Luca}, \citenamefont {Di~Castro}, \citenamefont {Kummer}
  \emph {et~al.}}]{Braicovich_2019arxiv}%
  \BibitemOpen
  \bibfield  {author} {\bibinfo {author} {\bibfnamefont {L.}~\bibnamefont
  {Braicovich}}, \bibinfo {author} {\bibfnamefont {M.}~\bibnamefont {Rossi}},
  \bibinfo {author} {\bibfnamefont {R.}~\bibnamefont {Fumagalli}}, \bibinfo
  {author} {\bibfnamefont {Y.}~\bibnamefont {Peng}}, \bibinfo {author}
  {\bibfnamefont {Y.}~\bibnamefont {Wang}}, \bibinfo {author} {\bibfnamefont
  {R.}~\bibnamefont {Arpaia}}, \bibinfo {author} {\bibfnamefont
  {D.}~\bibnamefont {Betto}}, \bibinfo {author} {\bibfnamefont {G.~M.}\
  \bibnamefont {De~Luca}}, \bibinfo {author} {\bibfnamefont {D.}~\bibnamefont
  {Di~Castro}}, \bibinfo {author} {\bibfnamefont {K.}~\bibnamefont {Kummer}},
  \emph {et~al.},\ }\href@noop {} {\bibfield  {journal} {\bibinfo  {journal}
  {arXiv preprint arXiv:1906.01270}\ } (\bibinfo {year} {2019})}\BibitemShut
  {NoStop}%
\bibitem [{\citenamefont {Braicovich}\ \emph {et~al.}(2010)\citenamefont
  {Braicovich}, \citenamefont {van~den Brink}, \citenamefont {Bisogni},
  \citenamefont {Sala}, \citenamefont {Ament}, \citenamefont {Brookes},
  \citenamefont {De~Luca}, \citenamefont {Salluzzo}, \citenamefont {Schmitt},
  \citenamefont {Strocov},\ and\ \citenamefont
  {Ghiringhelli}}]{Braicovich_2010_magnonRIXS}%
  \BibitemOpen
  \bibfield  {author} {\bibinfo {author} {\bibfnamefont {L.}~\bibnamefont
  {Braicovich}}, \bibinfo {author} {\bibfnamefont {J.}~\bibnamefont {van~den
  Brink}}, \bibinfo {author} {\bibfnamefont {V.}~\bibnamefont {Bisogni}},
  \bibinfo {author} {\bibfnamefont {M.~M.}\ \bibnamefont {Sala}}, \bibinfo
  {author} {\bibfnamefont {L.~J.~P.}\ \bibnamefont {Ament}}, \bibinfo {author}
  {\bibfnamefont {N.~B.}\ \bibnamefont {Brookes}}, \bibinfo {author}
  {\bibfnamefont {G.~M.}\ \bibnamefont {De~Luca}}, \bibinfo {author}
  {\bibfnamefont {M.}~\bibnamefont {Salluzzo}}, \bibinfo {author}
  {\bibfnamefont {T.}~\bibnamefont {Schmitt}}, \bibinfo {author} {\bibfnamefont
  {V.~N.}\ \bibnamefont {Strocov}}, \ and\ \bibinfo {author} {\bibfnamefont
  {G.}~\bibnamefont {Ghiringhelli}},\ }\href {\doibase
  10.1103/PhysRevLett.104.077002} {\bibfield  {journal} {\bibinfo  {journal}
  {Physical Review Letters}\ }\textbf {\bibinfo {volume} {104}},\ \bibinfo
  {pages} {077002} (\bibinfo {year} {2010})}\BibitemShut {NoStop}%
\bibitem [{\citenamefont {Tacon}\ \emph {et~al.}(2011)\citenamefont {Tacon},
  \citenamefont {Ghiringhelli}, \citenamefont {Chaloupka}, \citenamefont
  {Sala}, \citenamefont {Hinkov}, \citenamefont {Haverkort}, \citenamefont
  {Minola}, \citenamefont {Bakr}, \citenamefont {Zhou}, \citenamefont
  {Blanco-Canosa}, \citenamefont {Monney}, \citenamefont {Song}, \citenamefont
  {Sun}, \citenamefont {Lin}, \citenamefont {{De Luca}}, \citenamefont
  {Salluzzo}, \citenamefont {Khaliullin}, \citenamefont {Schmitt},
  \citenamefont {Braicovich},\ and\ \citenamefont
  {Keimer}}]{MLTparamagnonsNatPhys}%
  \BibitemOpen
  \bibfield  {author} {\bibinfo {author} {\bibfnamefont {M.~L.}\ \bibnamefont
  {Tacon}}, \bibinfo {author} {\bibfnamefont {G.}~\bibnamefont {Ghiringhelli}},
  \bibinfo {author} {\bibfnamefont {J.}~\bibnamefont {Chaloupka}}, \bibinfo
  {author} {\bibfnamefont {M.~M.}\ \bibnamefont {Sala}}, \bibinfo {author}
  {\bibfnamefont {V.}~\bibnamefont {Hinkov}}, \bibinfo {author} {\bibfnamefont
  {M.~W.}\ \bibnamefont {Haverkort}}, \bibinfo {author} {\bibfnamefont
  {M.}~\bibnamefont {Minola}}, \bibinfo {author} {\bibfnamefont
  {M.}~\bibnamefont {Bakr}}, \bibinfo {author} {\bibfnamefont {K.~J.}\
  \bibnamefont {Zhou}}, \bibinfo {author} {\bibfnamefont {S.}~\bibnamefont
  {Blanco-Canosa}}, \bibinfo {author} {\bibfnamefont {C.}~\bibnamefont
  {Monney}}, \bibinfo {author} {\bibfnamefont {Y.~T.}\ \bibnamefont {Song}},
  \bibinfo {author} {\bibfnamefont {G.~L.}\ \bibnamefont {Sun}}, \bibinfo
  {author} {\bibfnamefont {C.~T.}\ \bibnamefont {Lin}}, \bibinfo {author}
  {\bibfnamefont {G.~M.}\ \bibnamefont {{De Luca}}}, \bibinfo {author}
  {\bibfnamefont {M.}~\bibnamefont {Salluzzo}}, \bibinfo {author}
  {\bibfnamefont {G.}~\bibnamefont {Khaliullin}}, \bibinfo {author}
  {\bibfnamefont {T.}~\bibnamefont {Schmitt}}, \bibinfo {author} {\bibfnamefont
  {L.}~\bibnamefont {Braicovich}}, \ and\ \bibinfo {author} {\bibfnamefont
  {B.}~\bibnamefont {Keimer}},\ }\href {\doibase 10.1038/nphys2041} {\bibfield
  {journal} {\bibinfo  {journal} {Nature Physics}\ }\textbf {\bibinfo {volume}
  {7}},\ \bibinfo {pages} {725} (\bibinfo {year} {2011})}\BibitemShut {NoStop}%
\bibitem [{\citenamefont {Toth}\ and\ \citenamefont {Lake}(2015)}]{SpinW}%
  \BibitemOpen
  \bibfield  {author} {\bibinfo {author} {\bibfnamefont {S.}~\bibnamefont
  {Toth}}\ and\ \bibinfo {author} {\bibfnamefont {B.}~\bibnamefont {Lake}},\
  }\href {\doibase 10.1088/0953-8984/27/16/166002} {\bibfield  {journal}
  {\bibinfo  {journal} {Journal of Physics: Condensed Matter}\ }\textbf
  {\bibinfo {volume} {27}},\ \bibinfo {pages} {166002} (\bibinfo {year}
  {2015})}\BibitemShut {NoStop}%
\bibitem [{\citenamefont {Wang}\ \emph {et~al.}(2019)\citenamefont {Wang},
  \citenamefont {Zhou}, \citenamefont {Chen},\ and\ \citenamefont
  {Zhang}}]{Wang2019tj}%
  \BibitemOpen
  \bibfield  {author} {\bibinfo {author} {\bibfnamefont {Z.}~\bibnamefont
  {Wang}}, \bibinfo {author} {\bibfnamefont {S.}~\bibnamefont {Zhou}}, \bibinfo
  {author} {\bibfnamefont {W.-Q.}\ \bibnamefont {Chen}}, \ and\ \bibinfo
  {author} {\bibfnamefont {F.-C.}\ \bibnamefont {Zhang}},\ }\href@noop {} {\
  (\bibinfo {year} {2019})},\ \Eprint {http://arxiv.org/abs/1912.12581}
  {arXiv:1912.12581 [cond-mat.supr-con]} \BibitemShut {NoStop}%
\end{thebibliography}%
	
\end{document}